\documentclass[twocolumn,trackchanges]{aastex701}

\begin{document}

\title{A 4.5-s Quasiperiodic Spectral Oscillation in GRB 230307A: Evidence for Free Precession of a Post-Merger Magnetar?}

\author[orcid=0009-0009-2083-1999]{Run-Chao Chen}
\affiliation{School of Astronomy and Space Science, Nanjing University, Nanjing 210093, China}
\affiliation{Key Laboratory of Modern Astronomy and Astrophysics (Nanjing University), Ministry of Education, China}
\email[show]{chrczxx@smail.nju.edu.cn}

\author[0000-0002-5485-5042]{Jun Yang}
\affiliation{Institute for Astrophysics, School of Physics, Zhengzhou University, Zhengzhou 450001, China}
\email[hide]{jyang@smail.nju.edu.cn}

\author[0000-0003-4111-5958]{Bin-Bin Zhang}
\affiliation{School of Astronomy and Space Science, Nanjing University, Nanjing 210093, China}
\affiliation{Key Laboratory of Modern Astronomy and Astrophysics (Nanjing University), Ministry of Education, China}
\email[show]{bbzhang@nju.edu.cn}

\author[0009-0008-8053-2985]{Chen-Wei Wang}
\affiliation{State Key Laboratory of Particle Astrophysics, Institute of High Energy Physics, Chinese Academy of Sciences, Beijing 100049, China}
\affiliation{University of Chinese Academy of Sciences, Beijing 100049, China}
\email[hide]{cwwang@ihep.ac.cn}

\author[0009-0006-5506-5970]{Wen-Jun Tan}
\affiliation{State Key Laboratory of Particle Astrophysics, Institute of High Energy Physics, Chinese Academy of Sciences, Beijing 100049, China}
\affiliation{University of Chinese Academy of Sciences, Beijing 100049, China}
\email[hide]{tanwj@ihep.ac.cn}

\author[0000-0002-4771-7653]{Shao-Lin Xiong}
\affiliation{State Key Laboratory of Particle Astrophysics, Institute of High Energy Physics, Chinese Academy of Sciences, Beijing 100049, China}
\email[show]{xiongsl@ihep.ac.cn}

\author[0000-0002-9725-2524]{Bing Zhang}
\affiliation{Department of Physics, The University of Hong Kong, Pokfulam Road, Hong Kong, China}
\affiliation{The Hong Kong Institute for Astronomy and Astrophysics, The University of Hong Kong, Hong Kong, China}
\email[hide]{bzhang1@hku.hk}

\begin{abstract}
Millisecond magnetars, rapidly rotating neutron stars with ultra-strong magnetic fields, have long been proposed as central engines of gamma-ray bursts (GRBs). For GRBs produced by neutron star mergers, the survival of a long-lived magnetar remnant remains uncertain, as the merger remnant may rapidly collapse into a black hole. In GRB~230307A, multiwavelength observations together with a previously reported 909-Hz periodic signal consistent with millisecond spin in its prompt emission provide strong evidence that such a post-merger magnetar may power the burst.
Here we report the discovery of a quasiperiodic modulation with a characteristic period of 4.5~s in the spectral evolution of GRB~230307A, detected consistently across multiple gamma-ray instruments. The modulation is manifested as a coherent, energy-dependent variation of the spectral shape, with the strongest signature in the evolution of the peak energy.
Within the magnetar-engine framework, such a low-frequency modulation can be interpreted as a manifestation of large-scale periodic variations associated with the central engine. If interpreted in terms of free precession, the observed timescale implies a stellar ellipticity of $\epsilon \gtrsim 2.4 \times 10^{-4}$, corresponding to an internal magnetic field strength of $B_t \gtrsim 1.6 \times 10^{16}$~G, alongside a dipole field of $B_p \approx 5.6 \times 10^{15}$~G inferred from the early X-ray emission.
These results suggest that such systems may provide potential sources of post-merger gravitational waves (GWs), motivating targeted searches following GRB triggers.
\end{abstract}


\section{Introduction}\label{sec:intro}
The connection between neutron star (NS) mergers and GRBs was firmly established by the multimessenger event GW170817 \citep{2017PhRvL.119p1101A,2017ApJ...848L..13A,2017ApJ...848L..12A,2017Natur.551...71T}, which was accompanied by the short $\gamma$-ray burst GRB~170817A \citep{2017ApJ...848L..14G,2017ApJ...848L..15S,2018NatCo...9..447Z} and followed by the optical--infrared kilonova AT~2017gfo \citep{2017ApJ...848L..24V,2017Natur.551...67P,2017ApJ...848L..27T,2017ApJ...850L..37P}. This landmark observation provided a template for multimessenger studies of GRBs, demonstrating that the detection of a kilonova in the aftermath of a GRB can serve as a robust diagnostic of an NS-merger origin. 
At the same time, a key open question concerns the nature of the merger remnant, which may promptly collapse into a black hole or survive as a short- or long-lived NS \citep{2017ApJ...850L..19M,2018ApJ...852L..25R,2021GReGr..53...59S}. In the latter case, the remnant can act as a millisecond magnetar, injecting additional energy into the electromagnetic (EM) emission and shaping both the GRB and kilonova evolution \citep{2008MNRAS.385.1455M,2010ApJ...715..477Y,2013ApJ...776L..40Y,2014MNRAS.439.3916M}. Such EM signatures have been extensively modeled to constrain the lifetime and physical properties of post-merger magnetars \citep{2016ApJ...819...14S,2017ApJ...850L..19M,2018ApJ...861..114Y}.

Beyond broadband EM modeling, searches for periodic or quasiperiodic signals offer a complementary and potentially more direct probe of central-engine dynamics. Such timing signatures can trace characteristic timescales of the compact remnant and its environment, providing insights that are less dependent on detailed emission modeling \citep{1999ApJ...520..666P,2010A&A...516A..16L,2013PhRvD..87h4053S,2019ApJ...884L..16C,2023ApJ...947L..15M,2025MNRAS.540.2727L}. Motivated by these considerations, previous studies have searched for periodic features in the prompt emission of short GRBs, which are widely interpreted as originating from compact-object mergers \citep{2014ARA&A..52...43B}, as diagnostics of the progenitor system and central engine \citep{2013ApJ...777..132D,2002ApJ...576..932K,2023Natur.613..253C,2024JCAP...07..070L,2025MNRAS.537.2313Y,2026ApJ...998..289C}.

Recently, a small number of merger-driven GRBs with long-duration prompt emission have been identified, offering a new population in which timing diagnostics of the central engine can be explored. GRB~211211A, a minute-long burst at a luminosity distance of $\approx350$~Mpc with multiple emission episodes, showed no accompanying supernova but displayed convincing kilonova signatures, firmly linking this long-duration event to a NS merger and demonstrating that compact-object mergers can power GRBs with long prompt emission \citep{2022Natur.612..228T,2022Natur.612..223R,2022Natur.612..232Y}. Its multi-episode temporal structure, together with a well-defined X-ray plateau, indicates sustained energy injection from a long-lived magnetar remnant formed in the merger \citep{2022Natur.612..232Y}. GRB~230307A, another minute-long burst at $\approx300$~Mpc, exhibited a well-defined single-pulse prompt profile superposed with abundant fast variability \citep{2025ApJ...985..239Y}. It provides an even more compelling case: a spectroscopically confirmed kilonova was associated with this event \citep{2024Natur.626..737L,2024Natur.626..742Y}. Broadband observations further revealed a distinct soft X-ray component coexisting with the prompt $\gamma$-ray emission and forming an early plateau consistent with magnetar dipole spin-down, implying the presence of a long-lived magnetar central engine \citep{2025NSRev..12E.401S}. Moreover, a statistically significant 909-Hz periodic signal, consistent with the spin frequency of a millisecond magnetar, has been reported in its prompt $\gamma$-ray emission, consistent with the spin of a millisecond magnetar \citep{2025NatAs...9.1701C}. Given their long duration and high photon statistics, these bursts also enable sensitive searches for possible low-frequency quasiperiodic oscillation (QPO) signals \citep{2025A&A...702A.149H,2024ApJ...973L..33C,2024ApJ...970....6X,2024ApJ...967...26C}.

In this Letter, we report the discovery of a QPO with a characteristic period of about 4.5~s in GRB~230307A. Unlike previously reported high-frequency signals, this QPO appears at a much lower frequency and manifests as a coherent, energy-dependent modulation in the spectral evolution of the burst, producing strong periodic variations in both hardness ratios and time-resolved spectral parameters.
The structure of this paper is as follows:
Section~\ref{sec:obs} describes the observational data used in this work.
Section~\ref{sec:qpo} presents the observation and validation of the 4.5-s QPO in the spectral evolution of GRB~230307A.
Section~\ref{sec:dis} discusses the possible physical implications of this QPO, and Section~\ref{sec:concl}  summarizes our main conclusions.

\section{Observations}\label{sec:obs}

\subsection{Gamma-ray observation}
GRB~230307A triggered the Gravitational wave high-energy Electromagnetic Counterpart All-sky Monitor (GECAM) at 2023 March 7, 15:44:06.650~UT (hereafter $T_0$) \citep{2023GCN.33406....1X}. The \textit{Fermi} Gamma-ray Burst Monitor (GBM) was also triggered nearly simultaneously and independently confirmed the event \citep{2023GCN.33407....1D}. Owing to its extreme brightness, the burst was also detected by numerous other instruments, including \textit{Konus}-Wind \citep{2023GCN.33427....1S} and \textit{AstroSat} \citep{2023GCN.33415....1N,2023GCN.33437....1K}, as well as additional reports \citep{2023GCN.33410....1X,2023GCN.33412....1C,2023GCN.33418....1D,2023GCN.33424....1R,2023GCN.33438....1L,2023GCN.33478....1G,2023ApJ...956...97M}. The prompt emission exhibited multiple sharp spikes superposed on a broad envelope, with an energy-dependent duration of up to $\approx100$~s \citep{2025NSRev..12E.401S}.

In this work, we primarily use observations from GECAM-B \citep{2022RDTM....6...12L}, which provide the main dataset for our analysis. We select three gamma-ray detectors (GRDs) with the smallest incident angles, namely GRD01, GRD04, and GRD05, in order to maximize the effective area and signal quality. By combining the high- and low-gain readout modes, the effective energy coverage spans $\sim$22--6000~keV.

We further incorporate data from GECAM-C \citep{2023NIMPA105668586Z} and \textit{Fermi}/GBM \citep{2009ApJ...702..791M} as auxiliary datasets to verify the robustness of our results. For GECAM-C, we select the detector with the most favorable incident angle (GRD01), which provides a comparable energy coverage of $\sim$15--6000~keV. For \textit{Fermi}/GBM, we use the NaI detector na ($\sim$8--1000~keV) and the BGO detector b1 ($\sim$250~keV--40~MeV) with optimal viewing geometry. We note that the GBM data are affected by saturation during the brightest phase, leading to count losses due to buffer overflow and the resulting bad time intervals (BTIs) \citep{2023GCN.33551....1D}.

Together, these multi-instrument observations provide broad energy coverage and serve as cross-checks for both temporal and spectral analyses.

\subsection{Soft X-ray observation}
In the soft X-ray band, GRB~230307A was serendipitously observed by the Lobster Eye Imager for Astronomy (LEIA; 0.5--4~keV) \citep{2023GCN.33466....1L}, a wide-field focusing X-ray telescope that serves as the pathfinder of the Einstein Probe mission \citep{2022ApJ...941L...2Z,2025SCPMA..6839501Y}. The burst occurred about $0.6^{\circ}$ outside LEIA's nominal field of view; nevertheless, its extreme brightness enabled the cruciform arms of the instrument's point spread function to be clearly detected, yielding a significant photon count rate even without being in the focal spot. LEIA monitored the source from $T_{0}-94$~s to $T_{0}+667$~s, capturing a bright soft X-ray component that overlaps with the prompt gamma-ray emission and persists for several hundred seconds \citep{2025NSRev..12E.401S}.

\section{QPO in the spectral evolution of GRB~230307A}\label{sec:qpo}

Previous timing analyses of GRB~230307A have reported the presence of low-frequency QPOs in the prompt emission light curve. Using data from \textit{INTEGRAL}/SPI-ACS and \textit{Fermi}/GBM, \citet{2025A&A...702A.149H} identified a QPO at about 1.2~Hz with moderate significance, as well as a less significant signal at higher frequency. These results indicate that the prompt emission of GRB~230307A exhibits temporal quasiperiodicity.

Here we extend the investigation to the spectral domain and examine whether such quasiperiodic behavior is also reflected in the spectral evolution of the burst. Given the long duration of GRB~230307A and its exceptionally bright $\gamma$-ray emission, the burst provides a rare opportunity to probe spectral variability on short timescales with sufficient photon statistics.

\subsection{QPO in the Hardness Ratio Evolution}
The hardness ratio (HR) is defined as the ratio between the photon counts in two energy bands and serves as a model-independent proxy for spectral evolution. We define
\begin{equation}
\mathrm{HR}(t) = \frac{C_{\rm H}(t) - B_{\rm H}(t)}{C_{\rm S}(t) - B_{\rm S}(t)},
\end{equation}
where $C_{\rm H}$ and $C_{\rm S}$ are the observed counts in the hard and soft bands, and $B_{\rm H}$ and $B_{\rm S}$ are the corresponding background estimates.

We adopt a soft band of 22--50~keV and a hard band of 100--250~keV. The lower bound of 22~keV matches the effective sensitivity of GECAM-B for GRB~230307A. HR time series are constructed independently for GECAM-B and \textit{Fermi}/GBM over $[-100, 200]$~s relative to $T_0$, using a uniform bin width of 50~ms. This choice provides sufficient temporal resolution for frequencies up to 10~Hz while maintaining adequate photon statistics.

The burst interval is defined as $[-1, 99]$~s relative to $T_0$ \citep{2025NatAs...9.1701C}, and the background is modeled using a second-order polynomial fitted to the pre- and post-burst data. The fitted background is subtracted to obtain net light curves in each band. GECAM-C data are not used due to particle-induced contamination prior to the trigger, which introduces non-stationary background behavior \citep{2025NSRev..12E.401S}. Given the nearly identical detector design of GECAM-B and GECAM-C, this does not affect the robustness of our analysis. Due to the refined temporal resolution, some background-subtracted bins contain very low counts, leading to unstable HR values. We therefore exclude bins with net counts $\leq 1$ in either band, resulting in a non-uniformly sampled HR time series.

To search for transient periodicity in such data, we apply the weighted wavelet Z-transform (WWZ; \citealt{1996AJ....112.1709F}) to the $\log_{10}\mathrm{HR}$ time series within [$T_0-1$, $T_0+49$]~s\footnote{Data beyond $T_0+49$~s are excluded because the removal of low-count bins leaves insufficient valid data for a reliable periodicity search.}, using a sliding step of 0.25~s and a frequency resolution of 0.002~Hz. We use $\log_{10}\mathrm{HR}$ because it provides a more stable representation of relative spectral variations and is better suited for time--frequency analysis of a ratio-type quantity. All calculations are performed using the \texttt{libwwz} package.

\begin{figure*}
\centering
\includegraphics[width=0.8\linewidth]{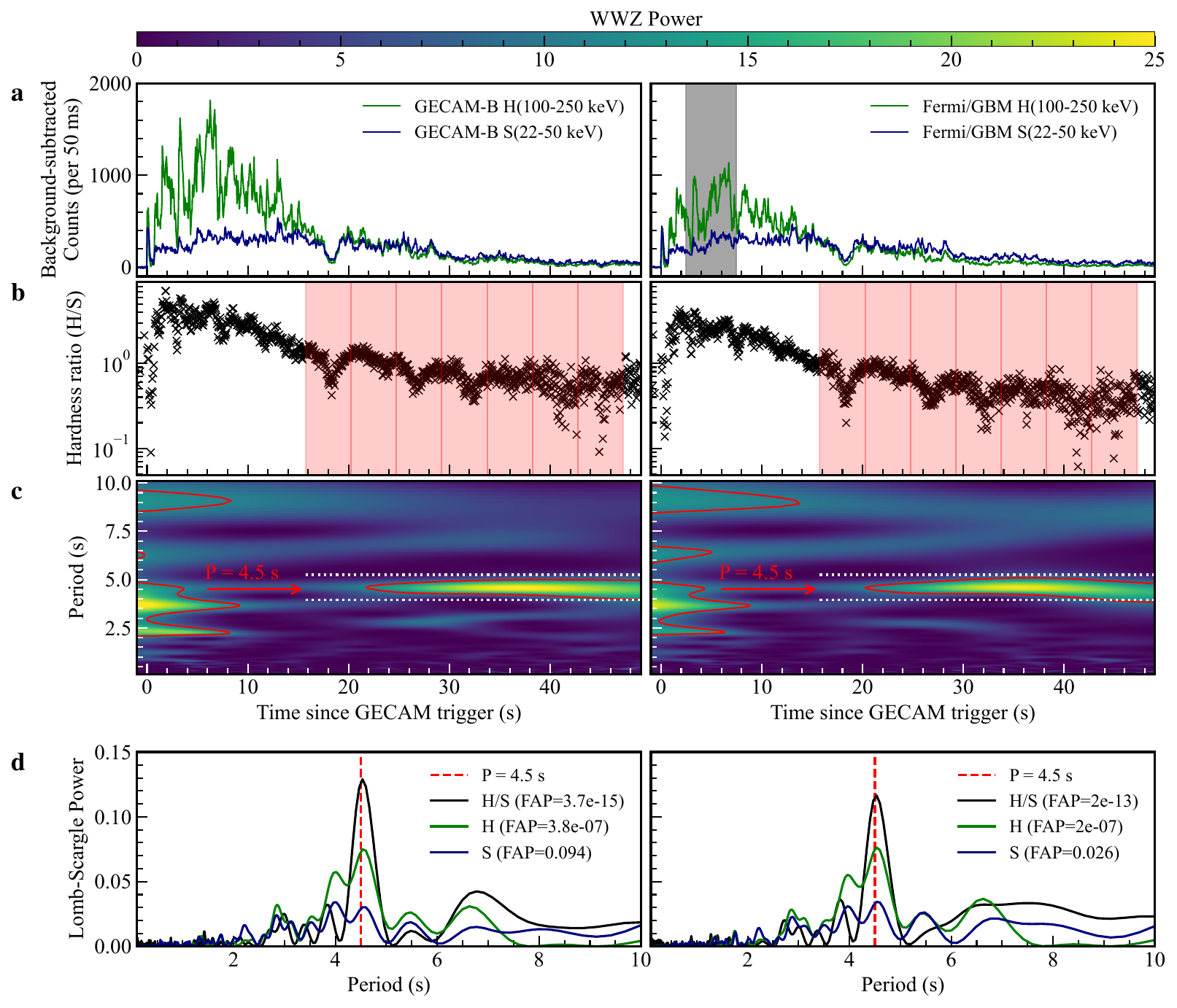}
\caption{\noindent\textbf{Detection of a 4.5-s QPO in the HR time series of GRB~230307A.}
\textbf{a,} Background-subtracted light curves from multiple instruments. The gray shaded region indicates the BTI in \textit{Fermi}/GBM \citep{2023GCN.33551....1D}.
\textbf{b,} HR time series. Black crosses represent HR measurements in 50-ms time bins; data points with net counts $\leq 1$ in either band are excluded. The red segments highlight seven consecutive cycles exhibiting quasiperiodic modulation.
\textbf{c,} Time--period spectrogram obtained from the WWZ of the $\log_{10}\mathrm{HR}$ time series, showing a prominent signal at 4.5~s (red arrow). The red contour marks the region enclosing the highest WWZ power (top 0.5\% of values), highlighting the dominant time--frequency structure associated with the modulation. The white dotted lines indicate the approximate period range set by the finite time-window resolution ($\approx$ 3.9--5.3~s), consistent with the quasi-stable extent of the detected feature.
\textbf{d,} LSPs of the $\log_{10}\mathrm{HR}$ time series and the background-subtracted light curves (soft and hard bands) within [$T_0 + 15.75$, $T_0 + 47.25$]~s. A consistent peak at 4.5~s is detected across all datasets and the labels indicate the single-trial FAP at the 4.5-s peak.}
\label{fig_HR}
\end{figure*}

Figure~\ref{fig_HR} presents the WWZ spectrogram and corresponding HR time series. A prominent quasiperiodic signal emerges after the brightest phase of GRB~230307A. Notably, an achromatic dip at around $T_0+18$~s, as a local extremum, provides a well-defined reference feature for characterizing the periodic modulation. In the absence of a priori confirmation of periodicity, we place this feature near the middle of a cycle rather than at the boundary of the analysis window, so that the local modulation pattern can be examined on both sides with reduced edge bias. Relative to this reference feature, the interval [$T_0+15.75$, $T_0+47.25$]~s covers approximately seven cycles of the 4.5-s modulation, consistently observed in both GECAM-B and \textit{Fermi}/GBM.

To assess the significance of this signal, we compute Lomb--Scargle periodograms (LSP; \citealt{2018ApJS..236...16V}) for the $\log_{10}\mathrm{HR}$ time series and the background-subtracted light curves in both energy bands over the same interval. The periodograms are evaluated over a frequency range of $0.01$--$10$~Hz using a uniform frequency grid oversampled by a factor of 10 relative to the Fourier frequency resolution $\Delta f = (7\times4.5~\mathrm{s})^{-1} \approx 0.03$~Hz set by the finite duration of the observed QPO signal. All datasets exhibit a consistent peak at a period of 4.5~s, with the strongest signal in the HR time series.

For this peak, we first estimate the single-trial false alarm probability (FAP) at the corresponding frequency\footnote{The single-trial FAP is estimated using a bootstrap resampling procedure with $10^4$ realizations \citep{2018ApJS..236...16V}. For very low FAP levels beyond the resolution of the bootstrap sampling, we adopt the analytic approximation of \citet{2008MNRAS.385.1279B}.}. We then account for multiple trials by multiplying the single-trial FAP by the total number of trials, given by the number of frequency bins in the oversampled grid ($\approx 3330$) and the number of temporal points within the interval identified from the WWZ spectrogram ($\approx 7\times4.5/0.25 \approx 126$), yielding a total of $\approx 4.2\times10^{5}$ trials.

Applying this correction to the measured single-trial FAP at the 4.5-s peak in the LSP of the HR time series (Figure~\ref{fig_HR}), we obtain trial-corrected FAP values of $\approx 1.57 \times 10^{-9}$ for GECAM-B and $\approx 8.39 \times 10^{-8}$ for \textit{Fermi}/GBM, corresponding to formal significances above the $5\sigma$ level. We note, however, that this correction likely overestimates the effective number of independent trials due to correlations in both the frequency grid and the time domain, and should therefore be regarded as a reference measure of the peak strength rather than a strict estimate of the statistical significance.

Furthermore, the single-trial FAP derived from the LSP at the 4.5-s peak is defined under the null hypothesis of independent, Gaussian (white-noise) fluctuations in the input time series. In contrast, GRB prompt emission is expected to exhibit temporally correlated variability (e.g., red-noise components), and the hardness ratio (HR), being a ratio of background-subtracted counts, does not strictly follow Poisson statistics. Constructing a physically motivated null hypothesis that captures these effects and their propagation into the HR is therefore non-trivial. Therefore, as a complementary approach, we perform a Monte Carlo (MC) test based on the observed count statistics to assess whether measurement uncertainties alone could produce a comparable modulation in the HR time series.

In each realization, the time grid and background level are fixed, while the observed counts in both bands are resampled assuming Poisson statistics. The same background fitting procedure, HR selection criteria, and LSP analysis are then applied. We record the maximum power near the target frequency $f_0 = 1/4.5~\mathrm{Hz}$ within a frequency window $\Delta f$, motivated by the finite number of observed cycles and the resulting spread of signal power over a narrow frequency range.

Using $10^5$ simulations, we obtain $p\approx8.79\times10^{-3}$ for GECAM-B and $p\approx6.81\times10^{-3}$ for \textit{Fermi}/GBM. As this test is performed directly on the observed data without assuming any specific model for the underlying variability, it evaluates the impact of counting statistics on the measured HR evolution. The results show that statistical fluctuations alone are unlikely to reproduce a modulation of the observed strength. Independently, the LSP analysis shows that the 4.5~s peak would be highly significant (at the $\gtrsim5\sigma$ level after trial correction) under the null hypothesis of no periodic signal. Taken together, these tests suggest that the observed 4.5~s peak is unlikely to be a statistical artifact and is consistent with a QPO signal.

\subsection{Validation of the QPO through Time-resolved Spectral Fitting}

The presence of a coherent QPO in HR evolution, its consistency across independent instruments, and the low probability of a statistical origin motivate a further investigation of its spectral origin based on refined time-resolved spectral analysis.

We perform time-resolved spectral fitting using combined data from GECAM-B, GECAM-C, and \textit{Fermi}/GBM. The spectra are divided into consecutive 0.15~s slices within the interval [$T_0+15.70$, $T_0+47.35$]~s. The starting time is intentionally offset from the timing analysis window [$T_0+15.75$, $T_0+47.25$]~s to avoid exact alignment between the spectral bins and the 4.5-s periodicity, thereby minimizing potential phase-locking effects and ensuring an independent test of the signal.

For GECAM-B, we use GRD01, GRD04, and GRD05, covering 40--350~keV (high-gain) and 700--6000~keV (low-gain). For GECAM-C, we include GRD01 in the 15--35 and 42--100~keV ranges, following the calibration procedure of \citet{2025NSRev..12E.401S}\footnote{Given the known particle-induced background variations in GECAM-C, its data are used only to supplement the low-energy coverage and do not dominate the spectral constraints.}. For \textit{Fermi}/GBM, we adopt the NaI detector na (10--30 and 40--900~keV) and the BGO detector b1 (300--38,000~keV), selected based on their favorable incident angles.

Each spectrum is fitted with a cutoff power-law (CPL) model,
\begin{equation}
N(E) = A \left(\frac{E}{100\,{\rm keV}}\right)^{\alpha} \exp\left(-\frac{E}{E_{\rm c}}\right),
\end{equation}
where $\alpha$ is the low-energy photon spectral index, $E_{\rm c}$ is the cutoff energy, and $A$ is the normalization. The peak energy of the $\nu f_{\nu}$ spectrum is given by $E_{\rm p} = (2+\alpha)E_{\rm c}$. This model is adopted because it provides the most stable fits across the entire time interval \citep{2025NSRev..12E.401S}, while models including a high-energy power-law component (e.g., Band function) yield poorly constrained $\beta$ values in low-count or late-time slices, limiting their usefulness for tracking temporal variability.

\begin{figure*}
\centering
\includegraphics[width=0.8\linewidth]{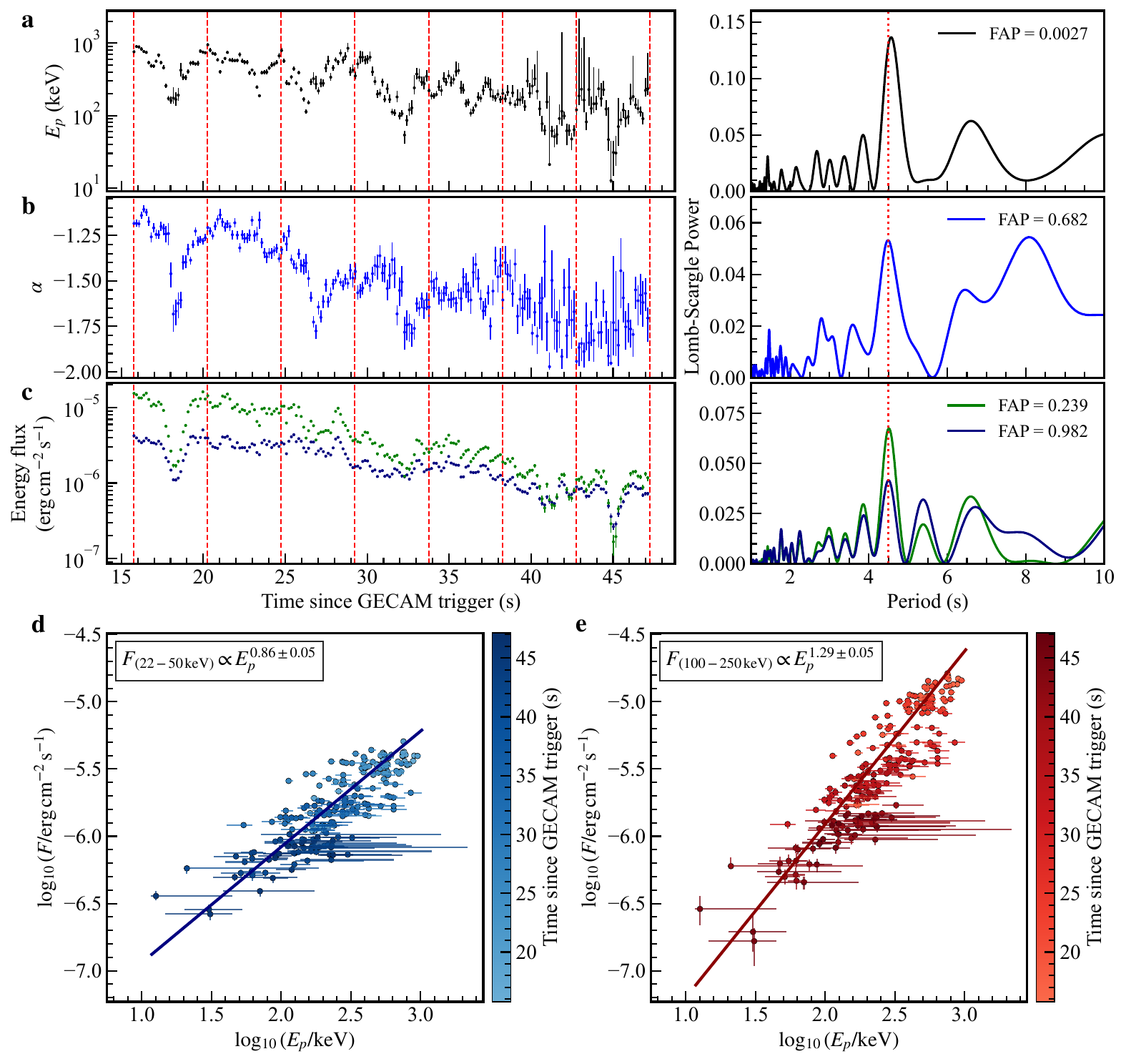}
\caption{\noindent\textbf{Spectral evolution and quasiperiodic behavior in GRB~230307A.}
\textbf{a,} Temporal evolution of the peak energy $E_{\rm p}$ (left) and its LSP (right), computed using $\log_{10}E_{\rm p}$.
\textbf{b,} Evolution of the low-energy photon spectral index $\alpha$ (left) and the corresponding LSP (right).
\textbf{c,} Energy flux ($F$) evolution in the soft (22--50~keV; blue) and hard (100--250~keV; green) bands (left), and their LSPs (right), computed using $\log_{10}F$.
Red dashed lines in the temporal panels mark the same seven cycles identified in the HR time series, while red dotted lines in the LSPs indicate the 4.5-s period. All spectral parameters are consistent with a 4.5-s QPO, with the strongest modulation observed in the evolution of $E_{\rm p}$.
\textbf{d,} Correlation between $\log_{10}E_{\rm p}$ and $\log_{10} F$ in the soft band (22--50~keV). The color scale indicates time evolution, and the solid line shows the best-fit power-law relation.
\textbf{e,} Same as panel \textbf{d}, but for the hard band (100--250~keV), showing a steeper dependence on $E_{\rm p}$.}
\label{fig_spec}
\end{figure*}

Spectral fitting is performed using the Bayesian package \texttt{bayspec} \citep{2022Natur.612..232Y,2023ApJ...947L..11Y,2025ApJ...989L..39Y}\footnote{\url{https://github.com/jyangch/bayspec}}, with the PGSTAT statistic \citep{1996ASPC..101...17A}. We track the evolution of the best-fit parameters $E_{\rm p}$, $\alpha$, and the energy flux integrated over selected bands in each 0.15~s slice, all results are presented in Table~\ref{sTab} and Figure~\ref{fig_spec}.

We then compute LSPs for these spectral quantities within [$T_0+15.75$, $T_0+47.25$]~s. According to the results presented in Figure~\ref{fig_spec}, all parameters consistently exhibit a quasiperiodic signal at 4.5~s, with the clearest modulation seen in $E_{\rm p}$ ($\text{FAP} \approx2\times10^{-3}$). This indicates that the periodicity identified in the HR is closely associated with intrinsic spectral evolution.

To further investigate the origin of the HR modulation, we examine the correlation between $E_{\rm p}$ and the energy flux $F$ in the soft (22--50~keV) and hard (100--250~keV) bands. In log--log space, we find power-law relations with indices of $\approx0.86$ for the soft band and $\approx1.29$ for the hard band (Figure~\ref{fig_spec}). This difference indicates that flux variations are more sensitive to changes in $E_{\rm p}$ at higher energies, where the spectrum lies closer to the peak of the $\nu f_{\nu}$ distribution. As a result, periodic oscillations in $E_{\rm p}$ naturally produce energy-dependent flux modulations, leading to a coherent oscillatory behavior in the hardness ratio.

These results support a physical interpretation of the HR QPO as a manifestation of periodic spectral evolution, rather than an artifact of counting statistics or background fluctuations.

\subsection{Phase–resolved Spectral Variability}

Having established the presence of the 4.5-s QPO in the spectral evolution, we next examine its energy dependence and its possible connection to the dip at $\simeq$18~s after $T_0$ in the light curve of GRB~230307A \citep{2025ApJ...985..239Y}. To this end, we perform a phase-resolved spectral variability analysis across multiple instruments, enabling a consistent comparison over a broad energy range and a direct assessment of phase coherence through folded profiles \citep[e.g.][]{2013Natur.500..312T}.

We use four independent data sets from GECAM-B, GECAM-C, and the \textit{Fermi}/GBM NaI and BGO detectors. For each instrument, the PHA channels are grouped into energy segments such that each segment contains at least $10^{4}$ photons within [$T_0+15.75$, $T_0+47.25$]~s. The segment boundaries $B_j$ are defined iteratively as
\begin{equation}
B_{j+1} = \min\left\{p>B_j ~\Big|~ N(B_j < {\rm PHA} < p) \geq 10^{4} \right\},
\end{equation}
where $N$ is the number of events in the specified channel range. This adaptive binning ensures comparable statistical quality across all energy segments.

We then fold the photon arrival times using $T_{\rm dip}=T_0 + 18$~s as the phase reference. The phase of each photon is computed as
\begin{equation}
\label{eq_pi}
\phi_i = \frac{t_i - T_{\rm dip}}{P} - \left\lfloor \frac{t_i - T_{\rm dip}}{P} \right\rfloor,
\end{equation}
where $P\simeq 4.5$~s and $\phi_i\in[0,1)$. Phase-folded profiles are constructed for each energy segment and instrument.

\begin{figure}
\centering
\includegraphics[width=\linewidth]{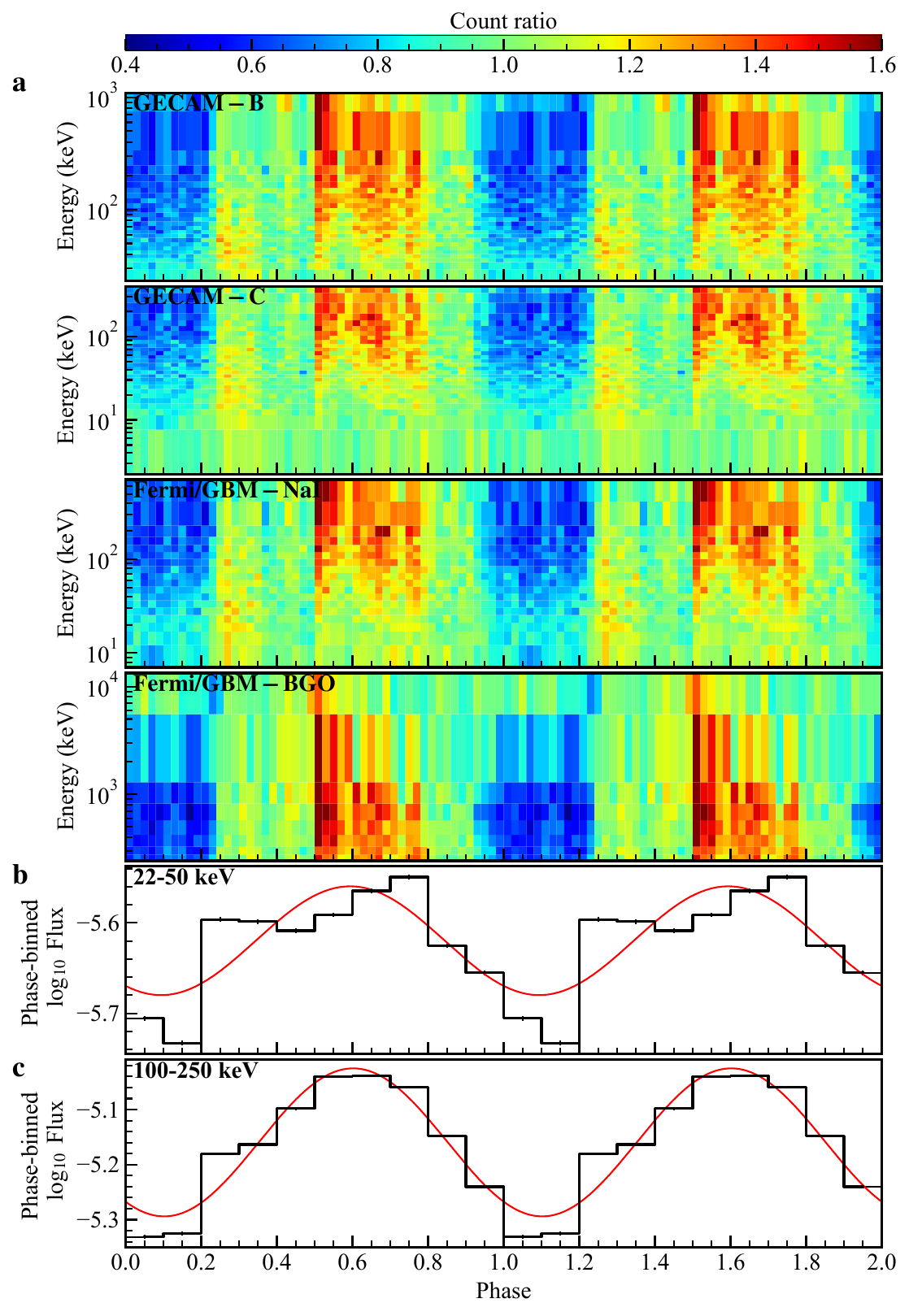}
\caption{\noindent\textbf{Phase--resolved spectral variability of GRB~230307A.}
\textbf{a} Phase--energy maps of GRB~230307A folded at a period of 4.5~s for GECAM-B, GECAM-C, \textit{Fermi}/GBM NaI, and BGO detectors, respectively. Photon arrival times are folded into 50 uniform phase bins, and the phase axis is displayed twice (0--2) to illustrate periodic continuity. The color scale represents the count ratio, defined as the photon counts in each phase bin normalized to the total counts within the corresponding energy segment. Since each segment contains approximately $10^{4}$ photons (except for the highest-energy bins), the count ratios are directly comparable across energies. The modulation amplitude decreases toward lower energies, while remaining phase-coherent and maintaining a stable pulse width up to the MeV range.
\textbf{b} Phase-folded flux profile in the soft band (22--50~keV), obtained from time-resolved spectral analysis. Black steps with error bars represent phase-binned measurements of $\log_{10} F$, and the red curve shows the best-fit sinusoidal model.
\textbf{c} Same as panel \textbf{b}, but for the hard band (100--250~keV). The modulation amplitude is significantly larger than in the soft band, while the phase remains aligned, consistent with the energy-dependent behavior seen in panel \textbf{a}.}
\label{fig_PS}
\end{figure}

The resulting phase--energy maps in Figure~\ref{fig_PS} show a clear energy dependence: the modulation amplitude decreases toward lower energies, while the phase remains coherent across the full energy range. This behavior is consistent with the previously reported energy-dependent periodicity at 909~Hz in GRB~230307A \citep{2025NatAs...9.1701C}, but is here revealed with improved clarity owing to the larger statistics and broader energy coverage.

To further quantify the 4.5-s modulation, we construct phase-binned flux profiles in representative soft (22--50~keV) and hard (100--250~keV) bands using the time-resolved spectral results. The fluxes are analyzed in logarithmic space and folded using Equation~\ref{eq_pi}. The phase interval is divided into 10 bins, within which weighted averages and uncertainties are computed.

The phase-folded profiles in Figure~\ref{fig_PS} exhibit a stable quasi-sinusoidal pattern over two cycles. We model the modulation as
\begin{equation}
\log_{10} F(\phi) = C + A \sin(2\pi \phi + \phi_0),
\end{equation}
where $C$ is the mean $\log F$, $A$ the modulation amplitude, and $\phi_0$ the phase offset. The best-fit amplitudes are $A_{\rm soft} = 0.060 \pm 0.001$ and $A_{\rm hard} = 0.134 \pm 0.001$, indicating that the hard-band modulation is more than twice as strong as in the soft band. The corresponding phase offsets, $\phi_{0,{\rm soft}} = 0.98 \pm 0.02$ and $\phi_{0,{\rm hard}} = 0.93 \pm 0.01$, are consistent with a nearly phase-aligned modulation across energies.

The fitted phase evolution is in good agreement with that observed in the phase--energy maps, with aligned peaks and troughs across all bands. This coherence suggests that the modulation is driven by a global spectral variation rather than independent fluctuations in different energy channels.

Notably, the reference time $T_{\rm dip}$ is located near phase zero of the modulation. Combined with the coherent phase evolution and the energy-dependent amplitude, this suggests that the observed dip may correspond to a particular phase of the underlying periodic modulation. In this picture, subsequent cycles would produce similar dip features at later times, but with progressively reduced contrast as the modulation amplitude decreases, making them less prominent in the light curve (Figure~\ref{fig_spec}).

Overall, the phase-resolved analysis shows that the 4.5-s signal is a coherent, energy-dependent modulation of the spectral shape, naturally linking the spectral QPO, flux variability, and the observed dip structure within a unified framework.

\section{Discussion}\label{sec:dis}
The 4.5-s QPO in GRB~230307A exhibits a clear energy dependence and phase coherence. To explore its possible physical origin, we consider two minimal and commonly invoked scenarios: geometrical modulation associated with Lense--Thirring precession of the accretion flow, and free precession of the central magnetar.

\subsection{Lense--Thirring precession}
One possible interpretation of the observed QPO is Lense--Thirring (LT) precession, a general relativistic effect caused by frame dragging around a rapidly rotating compact object \citep{1918PhyZ...19..156L}. In this scenario, a misalignment between the spin axis of the central remnant and the angular momentum axis of the accretion flow can induce precession of the inner disk or jet \citep{1998ApJ...492L..59S}.

To evaluate whether this mechanism can account for the observed 4.5-s modulation, we estimate the characteristic precession frequency near the inner disk edge. From the early soft X-ray observations by LEIA, assuming that the soft X-ray component is dominated by magnetic dipole radiation, we infer a dipole field strength of $B_p \approx 5.6\times10^{15}$~G for the central magnetar (Appendix~\ref{sec:A1}). We then estimate the Keplerian angular frequency at the Alfv\'en radius as
\begin{equation}
\Omega_K(r_A) = 9.5 \times 10^3\, B_{p,8}^{-6/7} R_6^{-18/7} \dot{M}_{17}^{3/7} M_0^{5/7}~{\rm rad\,s^{-1}},
\end{equation}
where $B_{p,8}=B_p/(10^8~{\rm G})$, $M_0=M/M_\odot$ is the stellar mass in solar units, $R_6=R/(10^6~{\rm cm})$ is the stellar radius in units of $10^6$~cm, and $\dot M_{17}=\dot M/(10^{17}~{\rm g\,s^{-1}})$ is the accretion rate in units of $10^{17}~{\rm g\,s^{-1}}$ \citep{1985Natur.316..239A}. Adopting $B_p=5.6\times10^{15}$~G, $M=2.37\,M_\odot$, $R=12$~km, and $\dot{M}\simeq0.1~M_\odot\,{\rm s^{-1}}$, we obtain
\begin{equation}
\Omega_K(r_A)\simeq 9.1\times10^3~{\rm rad\,s^{-1}},
\end{equation}
corresponding to a Keplerian frequency $\nu_K(r_A) \simeq 1.44\times10^3~{\rm Hz}$.

Using the characteristic inner-disk scale defined by $\nu_K(r_A)$ and adopting a spin frequency of $\nu_s=909$~Hz \citep{2025NatAs...9.1701C}, the LT precession frequency can be written as \citep{1998ApJ...492L..59S}
\begin{equation}
\nu_{\rm LT} = \frac{8\pi^2 I \nu_K^2 \nu_s}{c^2 M},
\end{equation}
where $I$ is the stellar moment of inertia, $M$ is the stellar mass, $c$ is the speed of light, and $\nu_K$ is the Keplerian frequency evaluated at the Alfv\'en radius $r_A$. 

To account for uncertainties in the neutron-star equation of state, we adopt a representative range $0.5 < I_{45}/M_0 < 2$, where $I_{45} \equiv I/(10^{45}~{\rm g\,cm^2})$ and $M_0 \equiv M/M_\odot$. Using the parameters above, we obtain
\begin{equation}
\nu_{\rm LT} \simeq 41\text{--}166~{\rm Hz},
\end{equation}
which is more than two orders of magnitude higher than the observed QPO frequency of $\approx0.22$~Hz ($P\simeq4.5$~s).

Therefore, the observed modulation is difficult to explain as local LT precession at the inner edge of the accretion flow in the magnetar framework. A lower-frequency global precession mode of the entire tilted flow may still be possible \citep[e.g.][]{2009MNRAS.397L.101I,2016MNRAS.461.1967I}. However, in merger-driven GRBs, the accretion flow is expected to transition into a rapidly evolving fallback-dominated regime on timescales of seconds, in which both the mass accretion rate and the characteristic disk radius change significantly with time. Such evolution would naturally lead to a noticeable temporal variation of the LT precession frequency, rather than the quasi-stable periodic signal observed over multiple cycles. In addition, the accretion flow in this phase is unlikely to maintain a coherent, long-lived tilted structure required for a well-defined precession mode. These considerations further disfavor a simple LT-precession interpretation of the observed 4.5-s QPO.

\subsection{Free Precession}

Given the extremely strong dipole magnetic field inferred for the central engine, a toroidal magnetic field of comparable or larger strength is expected to deform the neutron star away from axisymmetry \citep{2002PhRvD..66h4025C,2004ApJ...600..296I,2006A&A...450.1097B}. In the simplest biaxial rigid-body picture, such a deformation can induce free precession \citep{1974Natur.248..483P}, a mechanism that has also been discussed in the context of modulation observed in radio pulsars \citep[e.g.][]{2000Natur.406..484S}.

Under this simplified assumption, the stellar ellipticity can be estimated as
\begin{equation}
\epsilon \sim \frac{\nu_p}{\nu_s \cos\theta},
\label{eq_fp}
\end{equation}
where $\nu_p \simeq 0.22$~Hz is the precession frequency, $\nu_s = 909$~Hz is the spin frequency, and $\theta$ is the angle between the spin and precession axes. This yields $\epsilon \gtrsim 2.4 \times 10^{-4}$. Such a large ellipticity requires a strong internal toroidal magnetic field. Using the approximate scaling relation for magnetically deformed neutron stars\citep{2002PhRvD..66h4025C},
\begin{equation}
\epsilon \sim 10^{-4} \left(\frac{B_t}{10^{16}\,{\rm G}}\right)^2 ,
\end{equation}
we obtain $B_t \gtrsim 1.6 \times 10^{16}$~G. Although this exceeds the inferred dipole component, it remains plausible for a newly-formed post-merger magnetar \citep{2020ApJ...892L..34S,2021MNRAS.502.2482S}. These estimates suggest that the observed QPO is consistent with a free-precession interpretation of a magnetar possessing an extreme internal magnetic field.

A large ellipticity of this magnitude would also lead to significant GW emission. A non-axisymmetric rotating NS predominantly emits continuous GW near twice its spin frequency \citep{1979PhRvD..20..351Z}. The corresponding strain amplitude can be expressed as \citep{2015PASA...32...34L}
\begin{equation}
h_0(t) = \frac{4\pi^2 G}{c^4}\frac{I_{zz}\epsilon f_{\rm gw}^2(t)}{d},
\end{equation}
where $I_{zz}$ is the principal moment of inertia, $d$ is the source distance, and $f_{\rm gw}(t) \simeq 2\nu_s(t)$ is the GW frequency as the star spins down.

Given such a high ellipticity, both magnetic dipole radiation and GW emission contribute to the spin-down of the magnetar. The electromagnetic spin-down term is governed by the external dipole field $B_p$, while the GW component is determined by the stellar ellipticity (and hence the internal toroidal field $B_t$). The total energy loss rate is \citep{1967Natur.216..567P,2009MNRAS.398.1869D}
\footnote{We note that the magnetic-dipole term adopted here is intended as a simplified phenomenological parameterization. More realistic prescriptions can differ due to magnetospheric plasma effects \citep[e.g.,][]{2012ApJ...746L..24L}, and, in the case of newly-formed magnetars, additional uncertainties related to wind mass loading and magnetization may also affect the effective spin-down torque \citep[e.g.,][]{2011MNRAS.413.2031M}.}
\begin{equation}
\dot{E} = I\Omega\dot{\Omega} = -\frac{B_p^2 R^6 \Omega^4}{6c^3} - \frac{32GI^2 \epsilon^2 \Omega^6}{5c^5},
\label{eq:eloss}
\end{equation}
where the first term represents magnetic dipole losses and the second term corresponds to GW emission. Starting from the dipole spin-down interpretation of the soft X-ray light curve (Appendix~\ref{sec:A1}), we trace the expected spin evolution and assess the relative importance of the GW contribution.

\begin{figure}
\centering
\includegraphics[width=\linewidth]{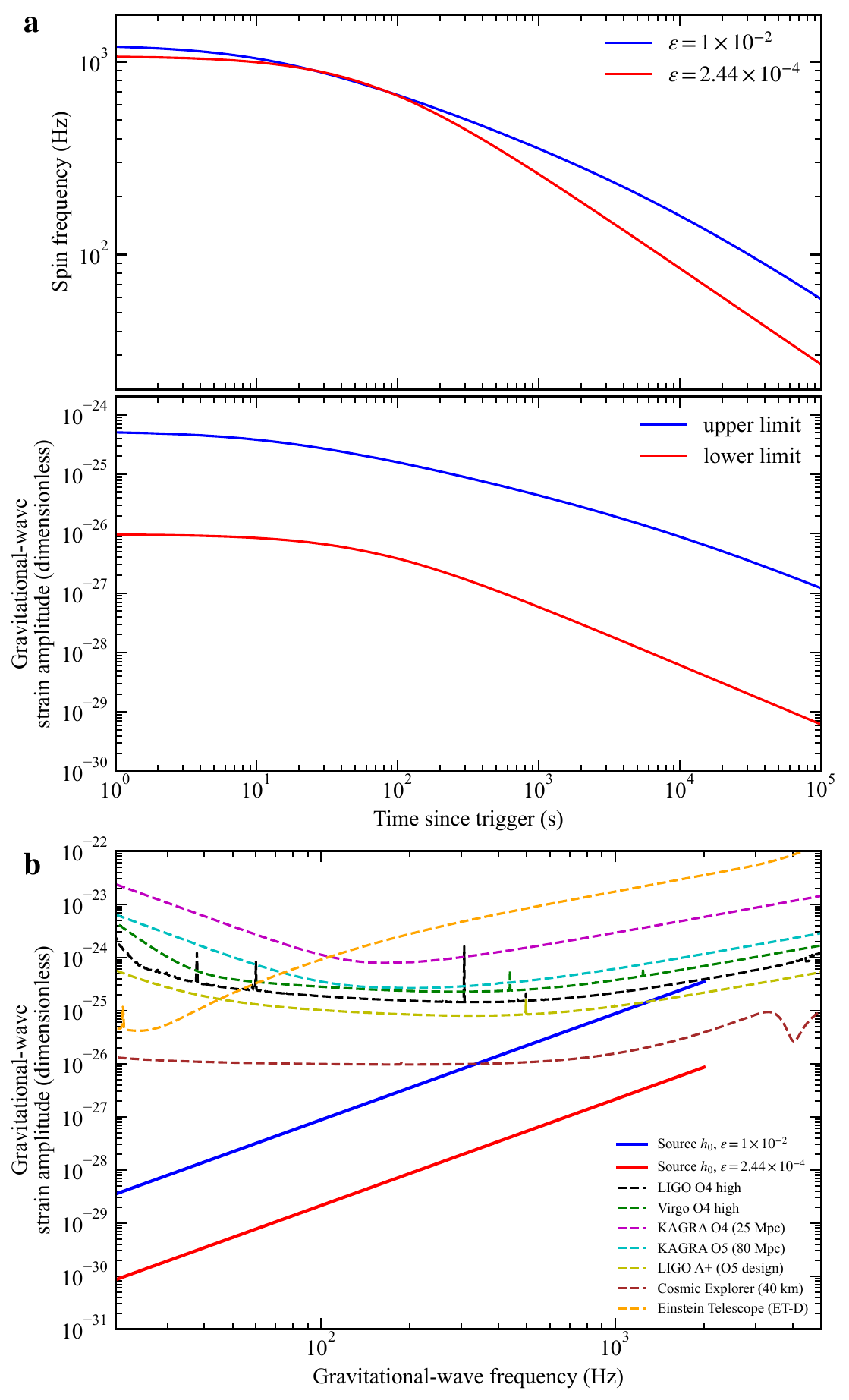}
\caption{\noindent\textbf{Evolution and detectability of gravitational-wave emission from the proto-magnetar in GRB~230307A.} 
\textbf{a,} Spin-frequency evolution of the magnetar for different assumed ellipticities, under the assumption that the soft X-ray emission is powered by magnetic dipole spin-down, together with the corresponding GW strain amplitudes as a function of time. 
\textbf{b,} Source GW strain as a function of frequency for the same ellipticities, compared with the effective sensitivity thresholds of current and planned detectors (LIGO/Virgo/KAGRA O4, LIGO A+, ET, and CE), estimated assuming an observation time of $10^5$~s and a detection threshold factor of $\Theta=15$. The comparison shows that, for the ellipticity inferred from the free-precession scenario, the signal amplitude approaches the design sensitivity of third-generation detectors.}
\label{fig_GW}
\end{figure}

Adopting a spectroscopic redshift of $z=0.0645$ for GRB~230307A \citep{2024Natur.626..737L}, we compute the GW strain evolution for ellipticities in the range $2.4\times10^{-4}$--$10^{-2}$, as shown in Figure~\ref{fig_GW}. We further compare the predicted strain as a function of frequency with the sensitivity curves of current and future ground-based GW detectors, including Advanced LIGO \citep{2015CQGra..32g4001L}, Advanced Virgo \citep{2015CQGra..32b4001A}, KAGRA \citep{2012CQGra..29l4007S}, the LIGO A+ design upgrade \citep{2020LRR....23....3A}, as well as next-generation observatories such as the Einstein Telescope (ET) and Cosmic Explorer (CE) \citep{2011CQGra..28i4013H,2021arXiv210909882E}. The detector thresholds are estimated by converting the noise power spectral densities $S_h(f)$ into effective strain amplitudes via
\begin{equation}
h_0^{\rm thr}(f) \simeq \Theta \sqrt{\frac{S_h(f)}{T_{\rm obs}}},
\end{equation}
where $T_{\rm obs}=10^{5}$~s is the assumed observation time and $\Theta=15$ is a representative detection threshold factor appropriate for targeted searches \citep{1998PhRvD..58f3001J,2007PhRvD..76h2001A}.

This comparison indicates that the expected GW signal from GRB~230307A is likely below the sensitivity of current detectors. Even under optimistic assumptions---including a well-constrained signal evolution based on the inferred source parameters and a long, coherent follow-up integration---the signal can approach the design sensitivity of third-generation detectors only if the ellipticity is sufficiently large.

Within the simplified free-precession framework, the periodic modulation observed in the EM emission provides a self-consistent set of magnetar parameters, including comparable toroidal and poloidal magnetic field strengths and a relatively large, yet physically allowable, ellipticity \citep{2007PhRvD..76d2001A,2008MNRAS.385..531H}. These parameters naturally lead to a predicted GW emission track in the strain–frequency plane, which places meaningful constraints on the detectability of post-merger GW signals. Although the prospects for directly tracking such an evolving signal remain uncertain, this scenario establishes a physically motivated connection between the observed QPO and the internal magnetic structure and rotational evolution of a post-merger magnetar. We note that the stability of mixed poloidal--toroidal magnetic configurations is itself a nontrivial issue. Analytic studies have shown that purely poloidal or purely toroidal fields are generally unstable, whereas mixed configurations can be stabilized only under specific conditions on the field geometry, energy partition, and internal stratification \citep[e.g.,][]{2013MNRAS.433.2445A}. Accordingly, the free-precession scenario adopted here should therefore be regarded as a simplified framework rather than a fully self-consistent description of the underlying magnetic-field configuration.

\section{Conclusions}\label{sec:concl}

We have identified a 4.5-s QPO in the spectral evolution of the prompt emission of GRB~230307A. This signal is consistently detected across independent instruments and manifests as a coherent, energy-dependent modulation in hardness ratios, time-resolved spectral parameters, and phase-resolved profiles, establishing it as an intrinsic feature of the burst.

To interpret its origin, we examined two representative scenarios within the context of a magnetar central engine. A simple LT precession at the inner accretion flow predicts characteristic frequencies in the range of tens to hundreds of Hz for the inferred source parameters, significantly exceeding the observed 0.22~Hz. This discrepancy disfavors a local disk-precession origin for the observed modulation. 

Within the magnetar-engine framework, the observed timescale instead points to large-scale periodic variations associated with the central object. If interpreted in terms of free precession, the modulation implies a stellar ellipticity of $\epsilon \gtrsim 2.4\times10^{-4}$, corresponding to an internal magnetic field strength of $B_t \gtrsim 1.6 \times 10^{16}$~G, in combination with a dipole field of $B_p \approx 5.6\times10^{15}$~G inferred from the early X-ray emission. These estimates provide indicative constraints on the physical conditions of a post-merger magnetar.

We further note that the WWZ power associated with the 4.5~s signal remains confined within the period range expected from the finite duration of the seven visible cycles, indicating that no strongly resolved secular drift is present over the $\approx31.5$~s interval (Figure~\ref{fig_HR}). Within the free-precession interpretation, the spin-evolution timescale is highly sensitive to the adopted spin-down parameters and can span from several to several tens of seconds for plausible millisecond spins and dipole magnetic fields of order $10^{15}$--$10^{16}$~G. Accordingly, free precession remains a viable interpretation within part of the parameter space, but is not uniquely favored by the current data; nor can the observed quasi-stability be used to place tight constraints on the spin-down properties.

A key issue in interpreting QPOs in GRBs is how (quasi)periodic variability originating from the central engine is transmitted through the jet and ultimately becomes observable in the prompt emission. For both the LT-precession and free-precession scenarios discussed above, this connection---manifested as geometric variations in the jet---remains poorly modeled, and no self-consistent physical framework currently exists to describe it quantitatively. Together with the previously reported 909-Hz signal \citep{2025NatAs...9.1701C}, the detection of this low-frequency QPO in GRB~230307A reveals a multi-scale timing structure in the prompt emission, suggesting that both rapid rotation and large-scale dynamical processes at the central engine may contribute to observable signatures across a wide range of timescales. The phase-resolved analysis further shows a similar energy-dependent behavior between the low- and high-frequency signals, hinting that they may arise from related physical processes that could potentially be understood within a unified framework, which can also account for the non-significant QPO signature during the brightest phase of the burst (see \citealt{2026ApJ...998..289C} for a detailed discussion). Establishing this connection, however, will require dedicated numerical modeling and simulations.

More broadly, such spectral QPOs offer a new avenue for probing the dynamics of GRB central engines and the physical processes governing their prompt emission. Future observations of similar events, particularly with improved photon statistics and broader energy coverage, will help to further clarify the origin and ubiquity of such spectral QPOs.

\begin{acknowledgments}
We acknowledge the support by the National Key Research and
Development Programs of China (2022YFF0711404), the National Natural Science Foundation of China (grant Nos. 12573046, 12121003), the Fundamental Research Funds for the Central Universities, and the Program for Innovative Talents and Entrepreneurs in Jiangsu. This work is also supported by the China Manned Space Program with grant No. CMS-CSST-2025-A17.
\end{acknowledgments}



\bibliography{ms}{}
\bibliographystyle{aasjournalv7}

\appendix

\section{Constraining the dipole field strength from the LEIA data}
\label{sec:A1}
To further constrain the dipole magnetic field strength of the millisecond magnetar, we make use of the LEIA X-ray light curve reported by \citet{2025NSRev..12E.401S}. The data points are taken directly from their published results, which were derived using a well-calibrated data reduction and background subtraction procedure. This approach ensures consistency with previous analyses while avoiding additional systematic uncertainties associated with independent data processing. The energy-loss formulation adopted for the X-ray modeling follows Equation~\ref{eq:eloss}, which relates the spin-down luminosity to the surface dipole magnetic field strength.

In this work, we introduce an additional constraint motivated by \citet{2025NatAs...9.1701C}, by assuming that the magnetar's spin frequency at $\tau=0$ (the onset of the spin-down evolution) was $\nu_0 = 909$~Hz. With this initial condition, the spin evolution is described as
\begin{equation}
\dot{E}_{\rm rot}(t) = -\eta \dot{E}_{\rm dip}(t),
\end{equation}
where $\eta$ is the electromagnetic radiation efficiency and $\dot{E}_{\rm dip}(t)$ is given by Equation~\ref{eq:eloss}. 

\begin{figure}[h]
\centering
\includegraphics[width=0.8\linewidth]{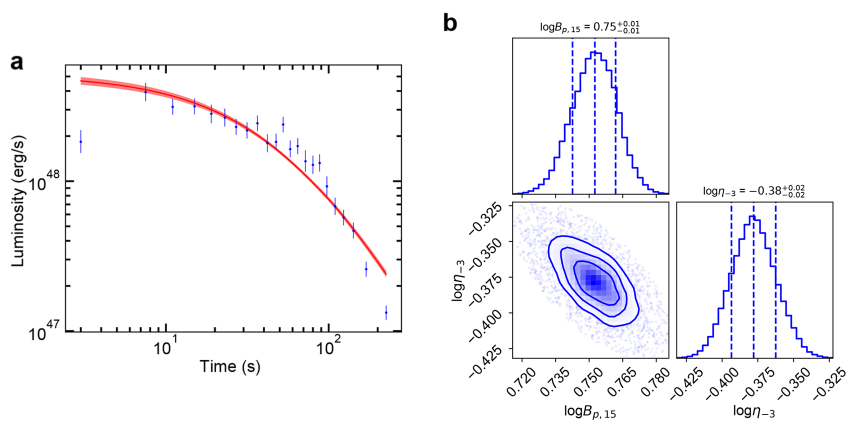}
\caption{\noindent\textbf{Constraints on the dipole magnetic field strength of the magnetar powering GRB~230307A.} 
\textbf{a,} The unabsorbed X-ray luminosity light curve in the 0.5--4~keV band \citep{2025NSRev..12E.401S}. Excluding the first data point, the light curve is fitted with a magnetar dipole spin-down model. The red line shows the best-fit model, and the shaded region indicates the $1\sigma$ confidence band.
\textbf{b,} Posterior distributions of the dipole magnetic field strength $B_{p,15}$ (in units of $10^{15}$~G) and the electromagnetic radiation efficiency $\eta_{-3}$ (in units of $10^{-3}$). Both parameters are sampled with log-flat priors, with $B_{p,15}$ in the range $[0,\,2]$ and $\eta_{-3}$ in the range $[-1,\,1]$ in logarithmic space. The blue dashed lines mark the best-fit values and $1\sigma$ uncertainties derived from the marginalized posterior distributions.}
\label{fig_A1}
\end{figure}

We performed parameter estimation using a Bayesian framework with the \texttt{MultiNest} sampler \citep{2009MNRAS.398.1601F}, adopting a likelihood function based on $\chi^2$ statistics. The number of live points was set to 1000 to ensure robust convergence of the posterior distributions. The LEIA data thereby allow us to simultaneously constrain $\eta$ and the dipole field strength $B_{\rm dip}$. The pairwise posterior distributions of these parameters are presented in Fig.~\ref{fig_A1}.

\clearpage
\startlongtable
\begin{deluxetable*}{cccccc}
\tablecaption{\textbf{Time resolved spectral fitting results and the corresponding fitting statistics.} Each time slice are marked as [$t_1$, $t_2$]~s from $T_0$. All errors represent the 1$\sigma$ uncertainties. \label{sTab}}
\tablehead{
\colhead{$t_1$ (s)} & 
\colhead{$t_2$ (s)} & 
\colhead{$\alpha$} & 
\colhead{${\rm log}E_{\rm p}$} &
\colhead{${\rm log}A$} &
\colhead{PGSTAT/d.o.f}
}
\startdata
$15.70$ & $15.85$ & $-1.18_{-0.02}^{+0.02}$ & $2.88_{-0.02}^{+0.02}$ & $-0.07_{-0.01}^{+0.01}$ & $1005.82/1189$ \\
$15.85$ & $16.00$ & $-1.18_{-0.02}^{+0.02}$ & $2.95_{-0.02}^{+0.02}$ & $-0.10_{-0.01}^{+0.01}$ & $1004.22/1189$ \\
$16.00$ & $16.15$ & $-1.19_{-0.02}^{+0.02}$ & $2.95_{-0.02}^{+0.02}$ & $-0.14_{-0.01}^{+0.01}$ & $961.63/1189$ \\
$16.15$ & $16.30$ & $-1.14_{-0.02}^{+0.02}$ & $2.92_{-0.02}^{+0.02}$ & $-0.14_{-0.01}^{+0.01}$ & $992.98/1189$ \\
$16.30$ & $16.45$ & $-1.11_{-0.02}^{+0.02}$ & $2.89_{-0.02}^{+0.02}$ & $-0.10_{-0.01}^{+0.01}$ & $1059.05/1189$ \\
$16.45$ & $16.60$ & $-1.12_{-0.02}^{+0.02}$ & $2.87_{-0.02}^{+0.02}$ & $-0.07_{-0.01}^{+0.01}$ & $964.01/1189$ \\
$16.60$ & $16.75$ & $-1.17_{-0.02}^{+0.03}$ & $2.73_{-0.02}^{+0.02}$ & $-0.12_{-0.01}^{+0.01}$ & $1017.92/1189$ \\
$16.75$ & $16.90$ & $-1.26_{-0.03}^{+0.03}$ & $2.69_{-0.03}^{+0.03}$ & $-0.19_{-0.01}^{+0.01}$ & $967.61/1189$ \\
$16.90$ & $17.05$ & $-1.21_{-0.02}^{+0.03}$ & $2.68_{-0.03}^{+0.02}$ & $-0.17_{-0.01}^{+0.01}$ & $979.93/1189$ \\
$17.05$ & $17.20$ & $-1.24_{-0.02}^{+0.03}$ & $2.75_{-0.02}^{+0.03}$ & $-0.20_{-0.01}^{+0.01}$ & $935.63/1189$ \\
$17.20$ & $17.35$ & $-1.25_{-0.02}^{+0.02}$ & $2.84_{-0.03}^{+0.02}$ & $-0.17_{-0.01}^{+0.01}$ & $980.31/1189$ \\
$17.35$ & $17.50$ & $-1.20_{-0.03}^{+0.03}$ & $2.73_{-0.03}^{+0.02}$ & $-0.18_{-0.01}^{+0.01}$ & $947.26/1189$ \\
$17.50$ & $17.65$ & $-1.21_{-0.04}^{+0.03}$ & $2.66_{-0.02}^{+0.03}$ & $-0.24_{-0.02}^{+0.01}$ & $965.22/1189$ \\
$17.65$ & $17.80$ & $-1.24_{-0.05}^{+0.04}$ & $2.41_{-0.03}^{+0.03}$ & $-0.36_{-0.03}^{+0.02}$ & $872.53/1189$ \\
$17.80$ & $17.95$ & $-1.24_{-0.06}^{+0.06}$ & $2.24_{-0.03}^{+0.03}$ & $-0.39_{-0.04}^{+0.03}$ & $903.32/1189$ \\
$17.95$ & $18.10$ & $-1.46_{-0.08}^{+0.06}$ & $2.21_{-0.04}^{+0.06}$ & $-0.62_{-0.05}^{+0.04}$ & $934.20/1189$ \\
$18.10$ & $18.25$ & $-1.68_{-0.10}^{+0.05}$ & $2.24_{-0.08}^{+0.14}$ & $-0.87_{-0.06}^{+0.03}$ & $843.13/1189$ \\
$18.25$ & $18.40$ & $-1.67_{-0.08}^{+0.06}$ & $2.22_{-0.09}^{+0.09}$ & $-0.86_{-0.05}^{+0.04}$ & $841.97/1189$ \\
$18.40$ & $18.55$ & $-1.62_{-0.08}^{+0.06}$ & $2.28_{-0.08}^{+0.10}$ & $-0.82_{-0.05}^{+0.04}$ & $823.89/1189$ \\
$18.55$ & $18.70$ & $-1.61_{-0.04}^{+0.05}$ & $2.63_{-0.08}^{+0.10}$ & $-0.74_{-0.02}^{+0.02}$ & $953.64/1189$ \\
$18.70$ & $18.85$ & $-1.40_{-0.04}^{+0.04}$ & $2.53_{-0.04}^{+0.05}$ & $-0.48_{-0.02}^{+0.02}$ & $893.15/1189$ \\
$18.85$ & $19.00$ & $-1.40_{-0.03}^{+0.04}$ & $2.52_{-0.04}^{+0.04}$ & $-0.40_{-0.02}^{+0.02}$ & $890.25/1189$ \\
$19.00$ & $19.15$ & $-1.33_{-0.03}^{+0.03}$ & $2.61_{-0.03}^{+0.03}$ & $-0.24_{-0.02}^{+0.01}$ & $950.00/1189$ \\
$19.15$ & $19.30$ & $-1.39_{-0.02}^{+0.02}$ & $2.72_{-0.03}^{+0.03}$ & $-0.18_{-0.01}^{+0.01}$ & $960.89/1189$ \\
$19.30$ & $19.45$ & $-1.30_{-0.02}^{+0.02}$ & $2.73_{-0.03}^{+0.02}$ & $-0.13_{-0.01}^{+0.01}$ & $1049.65/1189$ \\
$19.45$ & $19.60$ & $-1.23_{-0.02}^{+0.02}$ & $2.71_{-0.02}^{+0.03}$ & $-0.11_{-0.01}^{+0.01}$ & $1018.43/1189$ \\
$19.60$ & $19.75$ & $-1.28_{-0.03}^{+0.03}$ & $2.72_{-0.03}^{+0.03}$ & $-0.20_{-0.01}^{+0.01}$ & $1010.29/1189$ \\
$19.75$ & $19.90$ & $-1.27_{-0.02}^{+0.02}$ & $2.89_{-0.02}^{+0.03}$ & $-0.13_{-0.01}^{+0.01}$ & $967.07/1189$ \\
$19.90$ & $20.05$ & $-1.28_{-0.02}^{+0.02}$ & $2.88_{-0.02}^{+0.02}$ & $-0.04_{-0.01}^{+0.01}$ & $1031.91/1189$ \\
$20.05$ & $20.20$ & $-1.26_{-0.02}^{+0.02}$ & $2.87_{-0.03}^{+0.02}$ & $-0.14_{-0.01}^{+0.01}$ & $1007.22/1189$ \\
$20.20$ & $20.35$ & $-1.21_{-0.02}^{+0.02}$ & $2.98_{-0.02}^{+0.03}$ & $-0.11_{-0.01}^{+0.01}$ & $1044.82/1189$ \\
$20.35$ & $20.50$ & $-1.23_{-0.02}^{+0.02}$ & $2.90_{-0.02}^{+0.03}$ & $-0.19_{-0.01}^{+0.01}$ & $1018.71/1189$ \\
$20.50$ & $20.65$ & $-1.25_{-0.02}^{+0.02}$ & $2.86_{-0.03}^{+0.03}$ & $-0.24_{-0.01}^{+0.01}$ & $1013.18/1189$ \\
$20.65$ & $20.80$ & $-1.20_{-0.02}^{+0.03}$ & $2.85_{-0.03}^{+0.03}$ & $-0.23_{-0.01}^{+0.01}$ & $937.35/1189$ \\
$20.80$ & $20.95$ & $-1.16_{-0.03}^{+0.02}$ & $2.77_{-0.02}^{+0.03}$ & $-0.22_{-0.01}^{+0.01}$ & $944.22/1189$ \\
$20.95$ & $21.10$ & $-1.14_{-0.03}^{+0.03}$ & $2.61_{-0.02}^{+0.02}$ & $-0.14_{-0.01}^{+0.01}$ & $950.16/1189$ \\
$21.10$ & $21.25$ & $-1.18_{-0.02}^{+0.02}$ & $2.74_{-0.02}^{+0.02}$ & $-0.08_{-0.01}^{+0.01}$ & $899.40/1189$ \\
$21.25$ & $21.40$ & $-1.20_{-0.02}^{+0.02}$ & $2.79_{-0.03}^{+0.02}$ & $-0.10_{-0.01}^{+0.01}$ & $1009.95/1189$ \\
$21.40$ & $21.55$ & $-1.19_{-0.02}^{+0.02}$ & $2.87_{-0.02}^{+0.02}$ & $-0.08_{-0.01}^{+0.01}$ & $1012.21/1189$ \\
$21.55$ & $21.70$ & $-1.20_{-0.02}^{+0.02}$ & $2.77_{-0.02}^{+0.02}$ & $-0.12_{-0.01}^{+0.01}$ & $987.75/1189$ \\
$21.70$ & $21.85$ & $-1.25_{-0.03}^{+0.03}$ & $2.76_{-0.02}^{+0.03}$ & $-0.24_{-0.01}^{+0.01}$ & $988.09/1189$ \\
$21.85$ & $22.00$ & $-1.18_{-0.03}^{+0.02}$ & $2.76_{-0.02}^{+0.03}$ & $-0.22_{-0.01}^{+0.01}$ & $947.33/1189$ \\
$22.00$ & $22.15$ & $-1.25_{-0.02}^{+0.02}$ & $2.76_{-0.03}^{+0.02}$ & $-0.19_{-0.01}^{+0.01}$ & $962.43/1189$ \\
$22.15$ & $22.30$ & $-1.32_{-0.02}^{+0.02}$ & $2.84_{-0.03}^{+0.03}$ & $-0.29_{-0.01}^{+0.01}$ & $1042.32/1189$ \\
$22.30$ & $22.45$ & $-1.25_{-0.03}^{+0.03}$ & $2.75_{-0.03}^{+0.03}$ & $-0.25_{-0.01}^{+0.01}$ & $1002.21/1189$ \\
$22.45$ & $22.60$ & $-1.27_{-0.03}^{+0.03}$ & $2.74_{-0.03}^{+0.03}$ & $-0.26_{-0.01}^{+0.01}$ & $996.11/1189$ \\
$22.60$ & $22.75$ & $-1.27_{-0.03}^{+0.03}$ & $2.74_{-0.03}^{+0.03}$ & $-0.30_{-0.01}^{+0.01}$ & $900.99/1189$ \\
$22.75$ & $22.90$ & $-1.25_{-0.03}^{+0.03}$ & $2.59_{-0.03}^{+0.03}$ & $-0.24_{-0.02}^{+0.02}$ & $915.55/1189$ \\
$22.90$ & $23.05$ & $-1.17_{-0.03}^{+0.03}$ & $2.63_{-0.02}^{+0.02}$ & $-0.15_{-0.01}^{+0.01}$ & $938.03/1189$ \\
$23.05$ & $23.20$ & $-1.25_{-0.03}^{+0.03}$ & $2.58_{-0.03}^{+0.03}$ & $-0.21_{-0.02}^{+0.01}$ & $940.02/1189$ \\
$23.20$ & $23.35$ & $-1.24_{-0.04}^{+0.04}$ & $2.39_{-0.02}^{+0.03}$ & $-0.29_{-0.02}^{+0.02}$ & $914.95/1189$ \\
$23.35$ & $23.50$ & $-1.17_{-0.05}^{+0.05}$ & $2.27_{-0.02}^{+0.02}$ & $-0.26_{-0.03}^{+0.03}$ & $848.90/1189$ \\
$23.50$ & $23.65$ & $-1.29_{-0.03}^{+0.03}$ & $2.59_{-0.03}^{+0.03}$ & $-0.20_{-0.02}^{+0.01}$ & $986.24/1189$ \\
$23.65$ & $23.80$ & $-1.34_{-0.03}^{+0.02}$ & $2.60_{-0.03}^{+0.03}$ & $-0.19_{-0.01}^{+0.01}$ & $956.68/1189$ \\
$23.80$ & $23.95$ & $-1.37_{-0.03}^{+0.03}$ & $2.60_{-0.03}^{+0.04}$ & $-0.26_{-0.02}^{+0.01}$ & $1011.16/1189$ \\
$23.95$ & $24.10$ & $-1.35_{-0.03}^{+0.03}$ & $2.66_{-0.03}^{+0.04}$ & $-0.21_{-0.01}^{+0.01}$ & $989.27/1189$ \\
$24.10$ & $24.25$ & $-1.34_{-0.03}^{+0.02}$ & $2.70_{-0.03}^{+0.04}$ & $-0.29_{-0.02}^{+0.01}$ & $999.88/1189$ \\
$24.25$ & $24.40$ & $-1.33_{-0.03}^{+0.03}$ & $2.71_{-0.03}^{+0.04}$ & $-0.26_{-0.01}^{+0.01}$ & $974.09/1189$ \\
$24.40$ & $24.55$ & $-1.32_{-0.03}^{+0.03}$ & $2.73_{-0.03}^{+0.04}$ & $-0.23_{-0.01}^{+0.01}$ & $896.89/1189$ \\
$24.55$ & $24.70$ & $-1.38_{-0.02}^{+0.03}$ & $2.83_{-0.04}^{+0.04}$ & $-0.30_{-0.01}^{+0.01}$ & $1025.96/1189$ \\
$24.70$ & $24.85$ & $-1.33_{-0.03}^{+0.02}$ & $2.90_{-0.03}^{+0.03}$ & $-0.23_{-0.01}^{+0.01}$ & $960.35/1189$ \\
$24.85$ & $25.00$ & $-1.26_{-0.03}^{+0.03}$ & $2.75_{-0.02}^{+0.03}$ & $-0.21_{-0.01}^{+0.01}$ & $1002.15/1189$ \\
$25.00$ & $25.15$ & $-1.30_{-0.04}^{+0.04}$ & $2.45_{-0.03}^{+0.03}$ & $-0.34_{-0.02}^{+0.02}$ & $869.48/1189$ \\
$25.15$ & $25.30$ & $-1.23_{-0.03}^{+0.03}$ & $2.44_{-0.02}^{+0.03}$ & $-0.14_{-0.02}^{+0.01}$ & $957.77/1189$ \\
$25.30$ & $25.45$ & $-1.38_{-0.02}^{+0.03}$ & $2.53_{-0.03}^{+0.03}$ & $-0.16_{-0.01}^{+0.02}$ & $946.19/1189$ \\
$25.45$ & $25.60$ & $-1.41_{-0.03}^{+0.02}$ & $2.65_{-0.02}^{+0.03}$ & $-0.15_{-0.01}^{+0.01}$ & $1010.41/1189$ \\
$25.60$ & $25.75$ & $-1.43_{-0.03}^{+0.03}$ & $2.45_{-0.02}^{+0.04}$ & $-0.24_{-0.02}^{+0.01}$ & $965.61/1189$ \\
$25.75$ & $25.90$ & $-1.35_{-0.03}^{+0.03}$ & $2.59_{-0.02}^{+0.03}$ & $-0.17_{-0.01}^{+0.01}$ & $946.99/1189$ \\
$25.90$ & $26.05$ & $-1.41_{-0.04}^{+0.03}$ & $2.36_{-0.03}^{+0.03}$ & $-0.22_{-0.02}^{+0.02}$ & $831.26/1189$ \\
$26.05$ & $26.20$ & $-1.42_{-0.05}^{+0.04}$ & $2.23_{-0.03}^{+0.03}$ & $-0.30_{-0.03}^{+0.03}$ & $820.96/1189$ \\
$26.20$ & $26.35$ & $-1.40_{-0.06}^{+0.05}$ & $2.05_{-0.02}^{+0.03}$ & $-0.32_{-0.04}^{+0.03}$ & $851.72/1189$ \\
$26.35$ & $26.50$ & $-1.49_{-0.05}^{+0.04}$ & $2.18_{-0.03}^{+0.03}$ & $-0.34_{-0.03}^{+0.03}$ & $901.25/1189$ \\
$26.50$ & $26.65$ & $-1.59_{-0.02}^{+0.04}$ & $2.47_{-0.04}^{+0.04}$ & $-0.33_{-0.01}^{+0.02}$ & $1033.58/1189$ \\
$26.65$ & $26.80$ & $-1.73_{-0.03}^{+0.03}$ & $2.51_{-0.07}^{+0.09}$ & $-0.40_{-0.02}^{+0.02}$ & $1068.31/1189$ \\
$26.80$ & $26.95$ & $-1.78_{-0.04}^{+0.03}$ & $2.44_{-0.08}^{+0.11}$ & $-0.59_{-0.03}^{+0.02}$ & $868.36/1189$ \\
$26.95$ & $27.10$ & $-1.64_{-0.04}^{+0.03}$ & $2.34_{-0.05}^{+0.05}$ & $-0.47_{-0.02}^{+0.02}$ & $961.24/1189$ \\
$27.10$ & $27.25$ & $-1.70_{-0.03}^{+0.04}$ & $2.44_{-0.06}^{+0.06}$ & $-0.48_{-0.02}^{+0.02}$ & $888.15/1189$ \\
$27.25$ & $27.40$ & $-1.61_{-0.04}^{+0.03}$ & $2.45_{-0.04}^{+0.06}$ & $-0.47_{-0.02}^{+0.02}$ & $883.39/1189$ \\
$27.40$ & $27.55$ & $-1.52_{-0.04}^{+0.03}$ & $2.66_{-0.05}^{+0.06}$ & $-0.49_{-0.02}^{+0.02}$ & $891.03/1189$ \\
$27.55$ & $27.70$ & $-1.61_{-0.03}^{+0.03}$ & $2.75_{-0.06}^{+0.07}$ & $-0.50_{-0.02}^{+0.01}$ & $916.43/1189$ \\
$27.70$ & $27.85$ & $-1.57_{-0.04}^{+0.04}$ & $2.53_{-0.05}^{+0.05}$ & $-0.50_{-0.02}^{+0.02}$ & $868.72/1189$ \\
$27.85$ & $28.00$ & $-1.52_{-0.03}^{+0.03}$ & $2.66_{-0.04}^{+0.04}$ & $-0.38_{-0.01}^{+0.01}$ & $1000.89/1189$ \\
$28.00$ & $28.15$ & $-1.50_{-0.02}^{+0.02}$ & $2.77_{-0.04}^{+0.04}$ & $-0.26_{-0.01}^{+0.01}$ & $986.51/1189$ \\
$28.15$ & $28.30$ & $-1.46_{-0.02}^{+0.02}$ & $2.74_{-0.04}^{+0.03}$ & $-0.22_{-0.01}^{+0.01}$ & $936.92/1189$ \\
$28.30$ & $28.45$ & $-1.51_{-0.02}^{+0.02}$ & $2.86_{-0.04}^{+0.05}$ & $-0.31_{-0.01}^{+0.01}$ & $1018.61/1189$ \\
$28.45$ & $28.60$ & $-1.53_{-0.02}^{+0.03}$ & $2.77_{-0.05}^{+0.05}$ & $-0.40_{-0.01}^{+0.01}$ & $996.46/1189$ \\
$28.60$ & $28.75$ & $-1.49_{-0.03}^{+0.03}$ & $2.82_{-0.05}^{+0.06}$ & $-0.46_{-0.01}^{+0.01}$ & $980.57/1189$ \\
$28.75$ & $28.90$ & $-1.51_{-0.03}^{+0.03}$ & $2.93_{-0.05}^{+0.08}$ & $-0.53_{-0.01}^{+0.01}$ & $938.17/1189$ \\
$28.90$ & $29.05$ & $-1.44_{-0.05}^{+0.04}$ & $2.60_{-0.05}^{+0.07}$ & $-0.57_{-0.02}^{+0.02}$ & $925.74/1189$ \\
$29.05$ & $29.20$ & $-1.48_{-0.05}^{+0.03}$ & $2.63_{-0.04}^{+0.08}$ & $-0.63_{-0.03}^{+0.01}$ & $856.37/1189$ \\
$29.20$ & $29.35$ & $-1.45_{-0.04}^{+0.05}$ & $2.55_{-0.05}^{+0.06}$ & $-0.59_{-0.02}^{+0.02}$ & $847.31/1189$ \\
$29.35$ & $29.50$ & $-1.57_{-0.04}^{+0.04}$ & $2.71_{-0.07}^{+0.07}$ & $-0.65_{-0.02}^{+0.02}$ & $883.93/1189$ \\
$29.50$ & $29.65$ & $-1.55_{-0.04}^{+0.04}$ & $2.81_{-0.09}^{+0.09}$ & $-0.67_{-0.02}^{+0.02}$ & $892.43/1189$ \\
$29.65$ & $29.80$ & $-1.54_{-0.03}^{+0.04}$ & $2.81_{-0.07}^{+0.08}$ & $-0.58_{-0.02}^{+0.02}$ & $865.68/1189$ \\
$29.80$ & $29.95$ & $-1.48_{-0.04}^{+0.03}$ & $2.76_{-0.05}^{+0.06}$ & $-0.54_{-0.02}^{+0.01}$ & $878.58/1189$ \\
$29.95$ & $30.10$ & $-1.49_{-0.03}^{+0.04}$ & $2.78_{-0.07}^{+0.05}$ & $-0.55_{-0.02}^{+0.02}$ & $906.78/1189$ \\
$30.10$ & $30.25$ & $-1.53_{-0.03}^{+0.04}$ & $2.81_{-0.06}^{+0.08}$ & $-0.63_{-0.02}^{+0.02}$ & $921.57/1189$ \\
$30.25$ & $30.40$ & $-1.46_{-0.05}^{+0.04}$ & $2.60_{-0.05}^{+0.08}$ & $-0.66_{-0.03}^{+0.02}$ & $884.08/1189$ \\
$30.40$ & $30.55$ & $-1.42_{-0.06}^{+0.05}$ & $2.47_{-0.05}^{+0.07}$ & $-0.61_{-0.03}^{+0.03}$ & $878.14/1189$ \\
$30.55$ & $30.70$ & $-1.42_{-0.07}^{+0.07}$ & $2.30_{-0.05}^{+0.06}$ & $-0.69_{-0.04}^{+0.04}$ & $906.43/1189$ \\
$30.70$ & $30.85$ & $-1.36_{-0.07}^{+0.06}$ & $2.31_{-0.04}^{+0.07}$ & $-0.59_{-0.04}^{+0.03}$ & $816.13/1189$ \\
$30.85$ & $31.00$ & $-1.46_{-0.05}^{+0.05}$ & $2.47_{-0.06}^{+0.04}$ & $-0.64_{-0.02}^{+0.03}$ & $837.47/1189$ \\
$31.00$ & $31.15$ & $-1.42_{-0.09}^{+0.05}$ & $2.22_{-0.04}^{+0.06}$ & $-0.68_{-0.06}^{+0.03}$ & $831.93/1189$ \\
$31.15$ & $31.30$ & $-1.54_{-0.06}^{+0.05}$ & $2.37_{-0.06}^{+0.07}$ & $-0.74_{-0.03}^{+0.03}$ & $887.28/1189$ \\
$31.30$ & $31.45$ & $-1.47_{-0.07}^{+0.07}$ & $2.22_{-0.05}^{+0.07}$ & $-0.71_{-0.04}^{+0.04}$ & $914.76/1189$ \\
$31.45$ & $31.60$ & $-1.38_{-0.09}^{+0.09}$ & $2.05_{-0.04}^{+0.05}$ & $-0.60_{-0.06}^{+0.06}$ & $793.61/1189$ \\
$31.60$ & $31.75$ & $-1.46_{-0.08}^{+0.09}$ & $2.03_{-0.04}^{+0.05}$ & $-0.57_{-0.06}^{+0.06}$ & $922.11/1189$ \\
$31.75$ & $31.90$ & $-1.68_{-0.06}^{+0.05}$ & $2.26_{-0.07}^{+0.08}$ & $-0.74_{-0.04}^{+0.03}$ & $836.24/1189$ \\
$31.90$ & $32.05$ & $-1.55_{-0.09}^{+0.07}$ & $1.99_{-0.04}^{+0.05}$ & $-0.67_{-0.06}^{+0.05}$ & $920.39/1189$ \\
$32.05$ & $32.20$ & $-1.51_{-0.08}^{+0.09}$ & $2.03_{-0.04}^{+0.05}$ & $-0.69_{-0.06}^{+0.05}$ & $828.91/1189$ \\
$32.20$ & $32.35$ & $-1.78_{-0.10}^{+0.08}$ & $1.73_{-0.12}^{+0.06}$ & $-0.84_{-0.07}^{+0.07}$ & $799.22/1189$ \\
$32.35$ & $32.50$ & $-1.74_{-0.08}^{+0.08}$ & $1.93_{-0.07}^{+0.07}$ & $-0.82_{-0.06}^{+0.06}$ & $895.65/1189$ \\
$32.50$ & $32.65$ & $-1.74_{-0.10}^{+0.06}$ & $2.08_{-0.08}^{+0.10}$ & $-0.88_{-0.06}^{+0.03}$ & $851.44/1189$ \\
$32.65$ & $32.80$ & $-1.79_{-0.08}^{+0.05}$ & $2.11_{-0.09}^{+0.10}$ & $-0.83_{-0.05}^{+0.04}$ & $887.77/1189$ \\
$32.80$ & $32.95$ & $-1.76_{-0.06}^{+0.05}$ & $2.38_{-0.09}^{+0.17}$ & $-0.77_{-0.04}^{+0.03}$ & $856.51/1189$ \\
$32.95$ & $33.10$ & $-1.62_{-0.04}^{+0.03}$ & $2.57_{-0.07}^{+0.08}$ & $-0.60_{-0.02}^{+0.02}$ & $928.31/1189$ \\
$33.10$ & $33.25$ & $-1.65_{-0.05}^{+0.04}$ & $2.53_{-0.06}^{+0.12}$ & $-0.61_{-0.03}^{+0.02}$ & $980.47/1189$ \\
$33.25$ & $33.40$ & $-1.60_{-0.05}^{+0.03}$ & $2.46_{-0.05}^{+0.08}$ & $-0.63_{-0.03}^{+0.02}$ & $915.95/1189$ \\
$33.40$ & $33.55$ & $-1.65_{-0.06}^{+0.04}$ & $2.44_{-0.08}^{+0.10}$ & $-0.76_{-0.03}^{+0.03}$ & $871.82/1189$ \\
$33.55$ & $33.70$ & $-1.60_{-0.06}^{+0.04}$ & $2.54_{-0.07}^{+0.08}$ & $-0.71_{-0.03}^{+0.02}$ & $901.36/1189$ \\
$33.70$ & $33.85$ & $-1.64_{-0.06}^{+0.04}$ & $2.34_{-0.07}^{+0.08}$ & $-0.72_{-0.03}^{+0.02}$ & $837.58/1189$ \\
$33.85$ & $34.00$ & $-1.50_{-0.06}^{+0.06}$ & $2.27_{-0.05}^{+0.05}$ & $-0.62_{-0.03}^{+0.04}$ & $887.37/1189$ \\
$34.00$ & $34.15$ & $-1.54_{-0.05}^{+0.05}$ & $2.27_{-0.04}^{+0.05}$ & $-0.57_{-0.03}^{+0.03}$ & $824.19/1189$ \\
$34.15$ & $34.30$ & $-1.49_{-0.07}^{+0.04}$ & $2.29_{-0.03}^{+0.05}$ & $-0.55_{-0.04}^{+0.02}$ & $832.78/1189$ \\
$34.30$ & $34.45$ & $-1.57_{-0.06}^{+0.04}$ & $2.40_{-0.05}^{+0.09}$ & $-0.65_{-0.04}^{+0.02}$ & $836.54/1189$ \\
$34.45$ & $34.60$ & $-1.61_{-0.06}^{+0.06}$ & $2.35_{-0.07}^{+0.07}$ & $-0.72_{-0.04}^{+0.03}$ & $886.87/1189$ \\
$34.60$ & $34.75$ & $-1.57_{-0.05}^{+0.04}$ & $2.51_{-0.05}^{+0.09}$ & $-0.67_{-0.03}^{+0.02}$ & $875.16/1189$ \\
$34.75$ & $34.90$ & $-1.54_{-0.05}^{+0.04}$ & $2.46_{-0.05}^{+0.06}$ & $-0.57_{-0.03}^{+0.02}$ & $900.94/1189$ \\
$34.90$ & $35.05$ & $-1.60_{-0.04}^{+0.03}$ & $2.55_{-0.05}^{+0.08}$ & $-0.53_{-0.02}^{+0.02}$ & $780.57/1189$ \\
$35.05$ & $35.20$ & $-1.60_{-0.05}^{+0.04}$ & $2.45_{-0.05}^{+0.08}$ & $-0.62_{-0.03}^{+0.02}$ & $900.70/1189$ \\
$35.20$ & $35.35$ & $-1.60_{-0.04}^{+0.04}$ & $2.50_{-0.07}^{+0.07}$ & $-0.63_{-0.02}^{+0.02}$ & $909.53/1189$ \\
$35.35$ & $35.50$ & $-1.62_{-0.05}^{+0.04}$ & $2.61_{-0.06}^{+0.14}$ & $-0.66_{-0.03}^{+0.02}$ & $908.11/1189$ \\
$35.50$ & $35.65$ & $-1.55_{-0.05}^{+0.04}$ & $2.62_{-0.05}^{+0.11}$ & $-0.63_{-0.02}^{+0.02}$ & $870.03/1189$ \\
$35.65$ & $35.80$ & $-1.60_{-0.07}^{+0.06}$ & $2.38_{-0.07}^{+0.10}$ & $-0.80_{-0.04}^{+0.04}$ & $862.27/1189$ \\
$35.80$ & $35.95$ & $-1.48_{-0.09}^{+0.05}$ & $2.35_{-0.04}^{+0.09}$ & $-0.73_{-0.05}^{+0.03}$ & $866.28/1189$ \\
$35.95$ & $36.10$ & $-1.52_{-0.08}^{+0.06}$ & $2.16_{-0.05}^{+0.06}$ & $-0.60_{-0.05}^{+0.04}$ & $913.89/1189$ \\
$36.10$ & $36.25$ & $-1.56_{-0.06}^{+0.05}$ & $2.18_{-0.05}^{+0.05}$ & $-0.61_{-0.04}^{+0.04}$ & $914.47/1189$ \\
$36.25$ & $36.40$ & $-1.61_{-0.08}^{+0.07}$ & $2.03_{-0.04}^{+0.06}$ & $-0.71_{-0.05}^{+0.05}$ & $882.59/1189$ \\
$36.40$ & $36.55$ & $-1.45_{-0.08}^{+0.09}$ & $2.07_{-0.05}^{+0.05}$ & $-0.63_{-0.05}^{+0.06}$ & $845.71/1189$ \\
$36.55$ & $36.70$ & $-1.41_{-0.09}^{+0.06}$ & $2.14_{-0.04}^{+0.05}$ & $-0.61_{-0.05}^{+0.04}$ & $816.37/1189$ \\
$36.70$ & $36.85$ & $-1.49_{-0.08}^{+0.05}$ & $2.25_{-0.05}^{+0.07}$ & $-0.66_{-0.04}^{+0.03}$ & $828.95/1189$ \\
$36.85$ & $37.00$ & $-1.60_{-0.06}^{+0.05}$ & $2.24_{-0.05}^{+0.05}$ & $-0.64_{-0.03}^{+0.03}$ & $906.16/1189$ \\
$37.00$ & $37.15$ & $-1.67_{-0.05}^{+0.05}$ & $2.35_{-0.06}^{+0.08}$ & $-0.67_{-0.03}^{+0.03}$ & $863.26/1189$ \\
$37.15$ & $37.30$ & $-1.64_{-0.05}^{+0.05}$ & $2.46_{-0.07}^{+0.09}$ & $-0.68_{-0.03}^{+0.02}$ & $872.60/1189$ \\
$37.30$ & $37.45$ & $-1.57_{-0.06}^{+0.04}$ & $2.44_{-0.06}^{+0.09}$ & $-0.68_{-0.03}^{+0.02}$ & $882.93/1189$ \\
$37.45$ & $37.60$ & $-1.75_{-0.07}^{+0.05}$ & $2.36_{-0.11}^{+0.15}$ & $-0.85_{-0.04}^{+0.03}$ & $811.93/1189$ \\
$37.60$ & $37.75$ & $-1.56_{-0.08}^{+0.06}$ & $2.26_{-0.06}^{+0.10}$ & $-0.79_{-0.05}^{+0.04}$ & $877.30/1189$ \\
$37.75$ & $37.90$ & $-1.39_{-0.08}^{+0.07}$ & $2.24_{-0.05}^{+0.04}$ & $-0.71_{-0.04}^{+0.04}$ & $902.44/1189$ \\
$37.90$ & $38.05$ & $-1.48_{-0.09}^{+0.06}$ & $2.29_{-0.05}^{+0.06}$ & $-0.73_{-0.05}^{+0.03}$ & $856.77/1189$ \\
$38.05$ & $38.20$ & $-1.42_{-0.06}^{+0.10}$ & $2.22_{-0.05}^{+0.05}$ & $-0.66_{-0.04}^{+0.05}$ & $792.47/1189$ \\
$38.20$ & $38.35$ & $-1.60_{-0.06}^{+0.08}$ & $2.31_{-0.08}^{+0.08}$ & $-0.83_{-0.04}^{+0.05}$ & $821.71/1189$ \\
$38.35$ & $38.50$ & $-1.40_{-0.07}^{+0.09}$ & $2.23_{-0.05}^{+0.06}$ & $-0.78_{-0.04}^{+0.05}$ & $858.33/1189$ \\
$38.50$ & $38.65$ & $-1.46_{-0.09}^{+0.06}$ & $2.32_{-0.05}^{+0.10}$ & $-0.78_{-0.06}^{+0.03}$ & $828.64/1189$ \\
$38.65$ & $38.80$ & $-1.47_{-0.09}^{+0.09}$ & $2.17_{-0.04}^{+0.08}$ & $-0.74_{-0.06}^{+0.05}$ & $855.75/1189$ \\
$38.80$ & $38.95$ & $-1.57_{-0.07}^{+0.08}$ & $2.31_{-0.06}^{+0.10}$ & $-0.78_{-0.04}^{+0.05}$ & $868.22/1189$ \\
$38.95$ & $39.10$ & $-1.41_{-0.10}^{+0.08}$ & $2.15_{-0.05}^{+0.06}$ & $-0.77_{-0.06}^{+0.05}$ & $865.85/1189$ \\
$39.10$ & $39.25$ & $-1.69_{-0.10}^{+0.07}$ & $2.22_{-0.09}^{+0.14}$ & $-0.99_{-0.06}^{+0.04}$ & $851.56/1189$ \\
$39.25$ & $39.40$ & $-1.60_{-0.09}^{+0.09}$ & $2.22_{-0.08}^{+0.11}$ & $-0.96_{-0.06}^{+0.05}$ & $869.77/1189$ \\
$39.40$ & $39.55$ & $-1.65_{-0.10}^{+0.08}$ & $2.25_{-0.09}^{+0.13}$ & $-1.01_{-0.06}^{+0.05}$ & $799.55/1189$ \\
$39.55$ & $39.70$ & $-1.56_{-0.10}^{+0.10}$ & $2.12_{-0.06}^{+0.09}$ & $-0.95_{-0.07}^{+0.06}$ & $826.90/1189$ \\
$39.70$ & $39.85$ & $-1.60_{-0.12}^{+0.06}$ & $2.48_{-0.08}^{+0.32}$ & $-1.02_{-0.07}^{+0.03}$ & $818.98/1189$ \\
$39.85$ & $40.00$ & $-1.60_{-0.11}^{+0.05}$ & $2.31_{-0.07}^{+0.14}$ & $-0.94_{-0.06}^{+0.03}$ & $823.98/1189$ \\
$40.00$ & $40.15$ & $-1.79_{-0.07}^{+0.05}$ & $2.41_{-0.11}^{+0.23}$ & $-0.98_{-0.04}^{+0.03}$ & $839.47/1189$ \\
$40.15$ & $40.30$ & $-1.67_{-0.08}^{+0.06}$ & $2.51_{-0.10}^{+0.20}$ & $-0.93_{-0.04}^{+0.03}$ & $884.41/1189$ \\
$40.30$ & $40.45$ & $-1.80_{-0.10}^{+0.05}$ & $2.34_{-0.11}^{+0.56}$ & $-1.04_{-0.06}^{+0.03}$ & $861.42/1189$ \\
$40.45$ & $40.60$ & $-1.62_{-0.14}^{+0.11}$ & $2.09_{-0.10}^{+0.11}$ & $-1.07_{-0.09}^{+0.07}$ & $777.85/1189$ \\
$40.60$ & $40.75$ & $-1.60_{-0.23}^{+0.12}$ & $1.89_{-0.10}^{+0.11}$ & $-1.09_{-0.17}^{+0.09}$ & $829.03/1189$ \\
$40.75$ & $40.90$ & $-1.39_{-0.20}^{+0.20}$ & $1.79_{-0.06}^{+0.07}$ & $-0.95_{-0.16}^{+0.14}$ & $827.57/1189$ \\
$40.90$ & $41.05$ & $-1.72_{-0.16}^{+0.05}$ & $2.19_{-0.09}^{+0.41}$ & $-1.11_{-0.10}^{+0.03}$ & $767.34/1189$ \\
$41.05$ & $41.20$ & $-1.97_{-0.02}^{+0.08}$ & $1.32_{-0.00}^{+0.95}$ & $-1.29_{-0.05}^{+0.06}$ & $858.33/1189$ \\
$41.20$ & $41.35$ & $-1.50_{-0.21}^{+0.17}$ & $1.79_{-0.09}^{+0.07}$ & $-0.99_{-0.15}^{+0.12}$ & $763.92/1189$ \\
$41.35$ & $41.50$ & $-1.72_{-0.16}^{+0.17}$ & $1.71_{-0.14}^{+0.11}$ & $-1.15_{-0.12}^{+0.12}$ & $794.17/1189$ \\
$41.50$ & $41.65$ & $-1.78_{-0.12}^{+0.11}$ & $1.74_{-0.14}^{+0.13}$ & $-1.11_{-0.09}^{+0.07}$ & $815.17/1189$ \\
$41.65$ & $41.80$ & $-1.73_{-0.14}^{+0.12}$ & $1.96_{-0.08}^{+0.20}$ & $-0.95_{-0.12}^{+0.09}$ & $870.61/1189$ \\
$41.80$ & $41.95$ & $-1.92_{-0.07}^{+0.04}$ & $2.06_{-0.20}^{+1.09}$ & $-1.02_{-0.05}^{+0.02}$ & $812.28/1189$ \\
$41.95$ & $42.10$ & $-1.54_{-0.10}^{+0.11}$ & $1.99_{-0.05}^{+0.06}$ & $-0.83_{-0.07}^{+0.07}$ & $735.38/1189$ \\
$42.10$ & $42.25$ & $-1.80_{-0.13}^{+0.09}$ & $1.79_{-0.20}^{+0.10}$ & $-1.07_{-0.09}^{+0.07}$ & $818.62/1189$ \\
$42.25$ & $42.40$ & $-1.54_{-0.21}^{+0.16}$ & $1.79_{-0.07}^{+0.08}$ & $-0.92_{-0.17}^{+0.12}$ & $800.93/1189$ \\
$42.40$ & $42.55$ & $-1.72_{-0.16}^{+0.11}$ & $1.68_{-0.16}^{+0.09}$ & $-1.01_{-0.12}^{+0.08}$ & $787.40/1189$ \\
$42.55$ & $42.70$ & $-1.67_{-0.14}^{+0.12}$ & $1.80_{-0.10}^{+0.07}$ & $-0.97_{-0.10}^{+0.09}$ & $810.50/1189$ \\
$42.70$ & $42.85$ & $-1.94_{-0.04}^{+0.05}$ & $2.08_{-0.20}^{+0.75}$ & $-1.10_{-0.03}^{+0.03}$ & $876.96/1189$ \\
$42.85$ & $43.00$ & $-1.87_{-0.10}^{+0.03}$ & $2.27_{-0.13}^{+1.06}$ & $-1.10_{-0.06}^{+0.02}$ & $826.20/1189$ \\
$43.00$ & $43.15$ & $-1.84_{-0.10}^{+0.03}$ & $2.36_{-0.16}^{+0.57}$ & $-1.08_{-0.05}^{+0.03}$ & $779.18/1189$ \\
$43.15$ & $43.30$ & $-1.89_{-0.09}^{+0.05}$ & $2.21_{-0.17}^{+0.87}$ & $-1.14_{-0.05}^{+0.02}$ & $836.83/1189$ \\
$43.30$ & $43.45$ & $-1.75_{-0.14}^{+0.13}$ & $1.91_{-0.11}^{+0.13}$ & $-1.09_{-0.10}^{+0.08}$ & $794.45/1189$ \\
$43.45$ & $43.60$ & $-1.84_{-0.10}^{+0.06}$ & $2.36_{-0.22}^{+0.45}$ & $-1.18_{-0.05}^{+0.04}$ & $793.91/1189$ \\
$43.60$ & $43.75$ & $-1.74_{-0.12}^{+0.07}$ & $2.10_{-0.09}^{+0.17}$ & $-1.06_{-0.08}^{+0.04}$ & $837.18/1189$ \\
$43.75$ & $43.90$ & $-1.83_{-0.12}^{+0.06}$ & $2.16_{-0.13}^{+0.51}$ & $-1.08_{-0.07}^{+0.04}$ & $842.07/1189$ \\
$43.90$ & $44.05$ & $-1.61_{-0.10}^{+0.10}$ & $2.17_{-0.09}^{+0.11}$ & $-0.89_{-0.07}^{+0.07}$ & $829.12/1189$ \\
$44.05$ & $44.20$ & $-1.78_{-0.08}^{+0.05}$ & $2.37_{-0.11}^{+0.28}$ & $-1.01_{-0.05}^{+0.03}$ & $839.58/1189$ \\
$44.20$ & $44.35$ & $-1.78_{-0.09}^{+0.06}$ & $2.23_{-0.11}^{+0.22}$ & $-1.04_{-0.05}^{+0.04}$ & $889.32/1189$ \\
$44.35$ & $44.50$ & $-1.62_{-0.08}^{+0.07}$ & $2.36_{-0.09}^{+0.12}$ & $-0.97_{-0.05}^{+0.04}$ & $793.26/1189$ \\
$44.50$ & $44.65$ & $-1.72_{-0.12}^{+0.06}$ & $2.09_{-0.10}^{+0.12}$ & $-1.05_{-0.08}^{+0.04}$ & $908.66/1189$ \\
$44.65$ & $44.80$ & $-1.91_{-0.06}^{+0.15}$ & $1.67_{-0.15}^{+0.45}$ & $-1.32_{-0.06}^{+0.10}$ & $854.28/1189$ \\
$44.80$ & $44.95$ & $-1.95_{-0.01}^{+0.23}$ & $1.10_{-0.04}^{+0.55}$ & $-1.45_{-0.04}^{+0.21}$ & $783.43/1189$ \\
$44.95$ & $45.10$ & $-1.73_{-0.18}^{+0.37}$ & $1.49_{-0.33}^{+0.16}$ & $-1.41_{-0.15}^{+0.31}$ & $776.32/1189$ \\
$45.10$ & $45.25$ & $-1.77_{-0.14}^{+0.32}$ & $1.48_{-0.17}^{+0.24}$ & $-1.41_{-0.15}^{+0.25}$ & $689.69/1189$ \\
$45.25$ & $45.40$ & $-1.85_{-0.11}^{+0.14}$ & $1.85_{-0.26}^{+0.39}$ & $-1.40_{-0.09}^{+0.09}$ & $742.98/1189$ \\
$45.40$ & $45.55$ & $-1.44_{-0.21}^{+0.18}$ & $1.94_{-0.07}^{+0.11}$ & $-1.04_{-0.15}^{+0.12}$ & $791.81/1189$ \\
$45.55$ & $45.70$ & $-1.59_{-0.14}^{+0.14}$ & $2.08_{-0.11}^{+0.10}$ & $-1.10_{-0.09}^{+0.10}$ & $803.68/1189$ \\
$45.70$ & $45.85$ & $-1.75_{-0.13}^{+0.09}$ & $1.96_{-0.13}^{+0.11}$ & $-1.08_{-0.09}^{+0.06}$ & $867.43/1189$ \\
$45.85$ & $46.00$ & $-1.72_{-0.12}^{+0.10}$ & $1.99_{-0.11}^{+0.09}$ & $-1.04_{-0.08}^{+0.06}$ & $788.23/1189$ \\
$46.00$ & $46.15$ & $-1.76_{-0.10}^{+0.05}$ & $2.24_{-0.10}^{+0.20}$ & $-0.99_{-0.06}^{+0.03}$ & $831.57/1189$ \\
$46.15$ & $46.30$ & $-1.79_{-0.12}^{+0.03}$ & $2.26_{-0.07}^{+0.57}$ & $-0.98_{-0.07}^{+0.02}$ & $829.11/1189$ \\
$46.30$ & $46.45$ & $-1.55_{-0.10}^{+0.07}$ & $2.15_{-0.05}^{+0.09}$ & $-0.88_{-0.06}^{+0.05}$ & $826.23/1189$ \\
$46.45$ & $46.60$ & $-1.61_{-0.14}^{+0.08}$ & $2.10_{-0.07}^{+0.13}$ & $-0.93_{-0.09}^{+0.05}$ & $805.10/1189$ \\
$46.60$ & $46.75$ & $-1.58_{-0.11}^{+0.12}$ & $2.05_{-0.06}^{+0.07}$ & $-0.91_{-0.07}^{+0.08}$ & $802.14/1189$ \\
$46.75$ & $46.90$ & $-1.62_{-0.13}^{+0.12}$ & $1.96_{-0.08}^{+0.09}$ & $-1.01_{-0.09}^{+0.08}$ & $799.48/1189$ \\
$46.90$ & $47.05$ & $-1.57_{-0.10}^{+0.08}$ & $2.35_{-0.08}^{+0.17}$ & $-1.00_{-0.06}^{+0.04}$ & $795.17/1189$ \\
$47.05$ & $47.20$ & $-1.70_{-0.13}^{+0.06}$ & $2.37_{-0.09}^{+0.50}$ & $-1.07_{-0.08}^{+0.03}$ & $902.74/1189$ \\
$47.20$ & $47.35$ & $-1.72_{-0.06}^{+0.06}$ & $2.66_{-0.15}^{+0.27}$ & $-1.03_{-0.03}^{+0.03}$ & $862.25/1189$ \\
\enddata
\end{deluxetable*}

\end{document}